\documentclass[twocolumn,showpacs,prd,floatfix]{revtex4}

\usepackage{graphics,graphicx}

\newcommand{\beq}{\begin{equation}}
\newcommand{\eeq}{\end{equation}}
\newcommand{\bea}{\begin{eqnarray}}
\newcommand{\eea}{\end{eqnarray}}

\newcommand{\lie}{\mathcal{L}}

\begin{document}

\title{
Calibration of Moving Puncture Simulations}

\author{Bernd Br\"ugmann}
\affiliation{Theoretical Physics Institute, University of Jena, 07743 Jena, Germany}
\author{Jos\'e A. Gonz\'alez}
\affiliation{Theoretical Physics Institute, University of Jena, 07743 Jena, Germany}
\author{Mark Hannam}
\affiliation{Theoretical Physics Institute, University of Jena, 07743 Jena, Germany}
\author{Sascha Husa}
\affiliation{Theoretical Physics Institute, University of Jena, 07743 Jena, Germany}
\author{Ulrich Sperhake}
\affiliation{Theoretical Physics Institute, University of Jena, 07743 Jena, Germany}
\author{Wolfgang Tichy}
\affiliation{Department of Physics, Florida Atlantic University, Boca Raton,
  FL 33431, USA}

\date{\today}

\begin{abstract}
We present single and binary black hole simulations that follow the ``moving
puncture'' paradigm of simulating black-hole spacetimes without
excision, and use ``moving boxes'' mesh refinement. 
Focussing on binary black hole configurations
where the simulations cover roughly two orbits, 
we address five major issues determining the quality of our results: numerical
discretization error, finite extraction radius of the radiation
signal, physical appropriateness of initial data, gauge choice and
computational performance.
We also compare results we have obtained with the BAM code described here with the
independent LEAN code.
\end{abstract}


\maketitle


\section{Introduction}
\label{Introduction}

More than thirty years after the first numerical simulations of binary
black hole dynamics \cite{Hahn64,Smarr77}, the numerical relativity
community is now ready to compare binary black hole simulations with
experimental data. A series of recent breakthroughs
\cite{Pretorius:2005gq,Campanelli:2005dd,Baker05a,Bruegmann:2003aw,Scheel-etal-2006:dual-frame}  
has lead to a phase transition in the field: long-term evolutions of
inspiraling black holes that last for one orbit and more have been obtained
with several independent codes 
\cite{Pretorius:2006tp,Bruegmann:2003aw,Diener:2005mg,Campanelli:2006gf,Baker:2006yw,Herrmann:2006ks,Sperhake:2006cy,Scheel-etal-2006:dual-frame},  
and accurate gravitational-wave signals have been computed.

Coincidentally, these breakthroughs parallel the first science runs of
large-scale interferometric gravitational-wave 
observatories at design sensitivity \cite{Waldman06a}. The inspiral and merger
of black-hole binaries are considered to be among the most promising sources 
for this current generation of Earth-based gravitational wave detectors, and
it has become feasible for numerical relativity to contribute to the analysis
of experimental data. 

Such contributions will require large-scale parameter studies, and
correspondingly large computational resources. The crucial current technical
problem in the field is the efficiency of the
simulations, and the establishment of a ``data analysis pipeline'',
connecting analytical calculations of the early inspiral and late
ring-down phases with numerical simulations, and especially with
gravitational-wave searches in actual detector data.

In this paper we present a new version of the BAM code
\cite{Bruegmann96,Bruegmann97,Bruegmann:2003aw} for binary black hole 
simulations that follows the ``moving puncture'' paradigm of simulating
black-hole spacetimes without excision, and use ``moving boxes'' mesh
refinement. We give a detailed presentation of our methods, which will
serve as a reference for future work, and we use simple test cases of
single and orbiting black holes to calibrate our methods.

We give a detailed discussion of convergence and accuracy of our code,
and address further issues determining the quality of our results:
finite extraction radius of the radiation signal, physical
appropriateness of initial-data parameters, gauge choice and
computational performance.
We compare evolutions from different initial-data parameters and
conclude that Post-Newtonian methods provide excellent estimates for initial
positions, momenta and masses for quasi-circular orbits in the non-spinning
equal-mass case, removing the necessity of complex initial-data studies. We
present new results concerning the damping parameter $\eta$ in the popular
$\tilde\Gamma$-driver shift, which was originally introduced to handle long-term
coordinate drifts, but is now found to also have the beneficial effect of
magnifying the size of the apparent horizons and thus changing the spatial
resolution requirements. 
We also present timing and performance results: we have been able to
perform some of our highest resolution runs at a computational cost of
$\approx 500$ CPU-hours (see Table~\ref{tab:timing}), giving rise to the hope
that numerical relativity will indeed be capable of large-scale parameter
studies. 

In Sec.~\ref{sec:Basic} we recall the basic equations of the moving-puncture
method, followed by a detailed description of our wave extraction
algorithm and computation of ADM and Bondi quantities in Sec.~\ref{sec:asymptotics}. 
Our numerical methods and code structure are presented in
Sec.~\ref{sec:Numerics}. In Secs.~\ref{sec:Single}, 
~\ref{sec:Orbits} and ~\ref{sec:parameters} we describe our results 
for single black holes, orbiting black holes evolved using standard
quasi-circular-orbit initial-data parameters (allowing direct comparison with
the LEAN code \cite{Sperhake:2006cy}), and orbiting black holes with alternative 
initial-data parameters. We conclude with a discussion in Sec.~\ref{sec:Discussion}.

\section{The Puncture Method and Moving Punctures}\label{sec:Basic}

\subsection{Initial Data}\label{sec:ID}

We will model $N$-black-hole initial data by adopting the Brill-Lindquist
wormhole topology  \cite{Brill63} with $N+1$ asymptotically flat ends for our
initial geometry, thus enforcing the presence of N ``throats''.
The asymptotically flat ends are compactified and identified with points
$r_i$ on $R^3$. The coordinate singularities at the points $r_i$ resulting
from compactification are referred to as punctures. In the context of initial
data, punctures have been extensively studied and the treatment of the
singularity in the constraint equations is well understood
\cite{Dain:2002vm}. From a physical point of view the puncture representation of
black-hole initial data is particularly appealing because it provides a simple
prescription for associating masses, momenta and spins with any number of
black holes. 

Initial data consist of the positive-definite metric and extrinsic curvature
$(g_{ij}, K_{ij})$ induced on a spatial hypersurface $\Sigma$ with timelike
unit normal $n^i$. We choose our sign conventions as $n^i n_i = -1$, 
$$K_{ab} = -\frac{1}{2 \alpha} \lie_{n} g_{ab}.$$ We construct these data
using the conformal transverse-traceless decomposition of the initial-value
equations, outlined in \cite{York79}, and related to other conformal
decompositions in \cite{Pfeiffer:2002iy}. The spatial metric and intrinsic
curvature are conformally related to counterparts on a background space via an
initial conformal factor $\psi_0$, and the conformal extrinsic curvature is
split into trace and trace-free parts: 
\bea
	g_{ij} &=& \psi^4_0 \tilde{g}_{ij},
\\
	K_{ij} &=& \psi^{-2}_0 \bar A_{ij} + \frac{1}{3} g_{ij} K,
\eea
where $K = g^{ij} K_{ij}$ and $\bar{A}_{ij}$ is trace-free. 

We choose an initially flat background metric, $\tilde{g}_{ij} = \delta_{ij}$,
and a maximal slice, $K = 0$. The second choice decouples the Hamiltonian and momentum
constraints. The momentum constraint now takes the form \beq
\partial_j \bar A_{ij} = 0,
\eeq and admits the Bowen-York solutions \cite{Bowen80}
for any number of black holes with prescribed ADM linear and angular momenta. 

The Hamiltonian constraint becomes an elliptic equation for the
conformal factor with a solution of the form
\cite{Beig94,Beig:1994rp,Brandt97b,Husa-PhD,Dain01a,Dain:2002vm} 
\bea
	\psi_0 &=& \psi_{BL} + u,
\\
	\psi_{BL} &=& 1 + \sum_{i=1}^N \frac{m_i}{2 r_i}, \label{psiansatz}
\eea
The function $u$ is determined by an elliptic equation on
$R^3$ and is $C^\infty$ everywhere except at the punctures, where it
is $C^2$. The $m_i$ parameterize the mass of each black hole, but actually
equal the total mass of the black hole only in the special
case of the Schwarzschild spacetime. In the general case, the ADM mass at the
$i$th  asymptotically flat end (i.e., the puncture) is given by 
\beq
M_i = m_i \left(1 + u_i + \sum_{i \neq j} \frac{m_j}{2 d_{ij}} \right), \label{adm_punc}
\eeq 
where $u_i$ is the value of $u$ at the $i$th puncture, and $d_{ij}$ is
the coordinate distance between punctures $i$ and $j$. This quantity has been
found to agree within numerical uncertainty with the apparent-horizon mass $M_{AH,i}$ for
non-spinning punctures \cite{Tichy:2003qi}, and for spinning punctures we have
found it to agree with the black-hole mass given by a modification of the
Christodoulou formula \cite{Christodoulou70}, \beq
M_i^2 = (M_{AH,i})^2 + \frac{S_i^2}{4 (M_{AH,i})^2}. \label{BHmass}
\eeq 
Throughout this paper, lower-case $m_i$ will refer to the mass parameter
in the ansatz (\ref{psiansatz}), and $M_i$ will refer to the black-hole
mass given by Eq.~(\ref{adm_punc}). When we desire particular values of
$M_i$, we make initial guesses of $m_i$ by inverting Eq.~(\ref{adm_punc}),
and iteratively improve the $m_i$ based on successive values of $u_i$ until
the $M_i$ equal the desired values to within 0.02\%.
We will denote by $M$ the total black-hole mass of the system under investigation,
and typically use $M$ as the mass scale for quoting results (e.g.~when to express time or distance
in terms of a mass scale). 

To complete the definition of the initial data, we also need to specify
initial values for our gauge quantities, the lapse function $\alpha$ and shift
vector $\beta^i$. At time $t=0$ we use
\bea
	\alpha &=& 1 \quad \mbox{or}\quad \alpha = \psi_0^{-2},
\\
	\beta^i &=& 0. 
\eea
Both choices for the lapse have been used successfully, although
the ``pre-collapsed'' lapse suggested in 
\cite{Alcubierre02a,Campanelli:2005dd} is found to reduce
initial gauge dynamics and is our standard choice.
Both lapse and shift are updated by evolution equations depending on
the physical variables, as described below.

\subsection{Evolution System}

We evolve the initial data with the BSSN system
\cite{Shibata95,Baumgarte99}. On the initial slice the standard BSSN variables
$\phi$, $\tilde{g}_{ij}$, $\tilde{A}_{ij}$, $K$, and $\tilde{\Gamma}^i$ are
related to the variables in the conformal transverse-traceless decomposition by 
\bea
\phi             & = & \ln \psi_0 \\
\tilde{A}_{ij}   & = & \psi^{-6} \bar{A}_{ij} \\
\tilde{\Gamma}^i & = & - \partial_j \tilde{g}^{ij},
\eea
and $\tilde{g}_{ij}$ and $K$ are unchanged. These variables are evolved using 
\bea
\label{phidot}
	\partial_0 \phi &=& - \frac{1}{6}\alpha K,
\\
\label{gdot}
	\partial_0 \tilde g_{ij} &=& -2\alpha\tilde A_{ij},
\\
	\partial_0 \tilde A_{ij} &=& e^{-4\phi}
	[-D_iD_j\alpha + \alpha R_{ij}]^{TF}
\nonumber
\\
\label{Adot}
	&& + \alpha(K\tilde A_{ij} - 2 \tilde A_{ik}\tilde {A^k}_j),
\\
\label{Kdot}
	\partial_0 K &=& -D^iD_i\alpha + 
	\alpha(\tilde A_{ij}\tilde A^{ij} + \frac{1}{3} K^2),
\\
\label{Gdot}
	\partial_t \tilde\Gamma^i &=& \tilde{g}^{ij} \partial_j \partial_k
        \beta^i + \frac{1}{3} \tilde{g}^{ij} \partial_j \partial_k \beta^k +
        \beta^j \partial_j \tilde{\Gamma}^i  \nonumber \\ 
	&&  - \tilde{\Gamma}^j \partial_j \beta^i + \frac{2}{3}
        \tilde{\Gamma}^{i} \partial_j \beta^j - 2 \tilde{A}^{ij} \partial_j
        \alpha \nonumber  \\ 
	&& + 2 \alpha \left (\tilde{\Gamma}^i_{jk} \tilde{A}^{jk} + 6
          \tilde{A}^{ij} \partial_j \phi - \frac{2}{3} \tilde{g}^{ij}
          \partial_j K \right) , 
\eea
where $\partial_0 = \partial_t - \lie_\beta$, $D_i$ is the covariant
derivative with respect to the conformal metric $\tilde{g}_{ij}$, and ``TF''
denotes the trace-free part of the expression with respect to the {\it
  physical} metric, $X_{ij}^{TF} = X_{ij} - \frac{1}{3} g_{ij} X_k^k$. The
Ricci tensor $R_{ij}$ is given by \bea 
R_{ij} & = & \tilde{R}_{ij} + R^{\phi}_{ij} \\
\tilde{R}_{ij} & = &   - \frac{1}{2} \tilde{g}^{lm} \partial_l \partial_m
\tilde{g}_{ij} + \tilde{g}_{k(i} \partial_{j)} \tilde{\Gamma}^k +
\tilde{\Gamma}^k \tilde{\Gamma}_{(ij)k} +        \nonumber   \\ 
                       &  & \tilde{g}^{lm} \left( 2 \tilde{\Gamma}^k_{l(i}
                         \tilde{\Gamma}_{j)jm} + \tilde{\Gamma}^k_{im}
                         \tilde{\Gamma}_{klj} \right), \\ 
R^{\phi}_{ij}  & = & - 2 D_i D_j \phi - 2 \tilde{g}_{ij} D^k D_k \phi + 4 D_i
\phi D_j \phi - \nonumber \\ 
                        &    & 4 \tilde{g}_{ij} D^k \phi D_k \phi.
\eea 
The Lie derivatives of the tensor densities $\phi$, $\tilde{g}_{ij}$ and
$\tilde{A}_{ij}$ (with weights $1/6$, $-2/3$ and $-2/3$) are 
\begin{eqnarray*}
\lie_\beta \phi & = & \beta^k \partial_k \phi + \frac{1}{6} \partial_k \beta^k, \\
\lie_\beta \tilde{g}_{ij} & = & \tilde{g}_{ij} \partial_k \tilde{g}_{ij} +
\tilde{g}_{ik} \partial_j \beta^k + \tilde{g}_{jk} \partial_i \beta^k -
\frac{2}{3} \tilde{g}_{ij} \partial_k \beta^k, \\ 
\lie_\beta \tilde{A}_{ij} & = &  \tilde{A}_{ij} \partial_k \tilde{A}_{ij} +
\tilde{A}_{ik} \partial_j \beta^k + \tilde{A}_{jk} \partial_i \beta^k -
\frac{2}{3} \tilde{A}_{ij} \partial_k \beta^k. 
\end{eqnarray*}

It is common to evolve the BSSN system as a partially constrained scheme,
where one or both of the algebraic constraints  $det(g) = 1$ and 
$Tr(A_{ij})=0$ are enforced at every full or intermediate time step of the
evolution scheme. 
This has been found to be necessary to obtain stable, accurate evolutions of
black-hole punctures in several cases, see e.g.,
\cite{Alcubierre99d,Alcubierre02a}. Likewise, the algebraic constraints and
the first-order differential constraint   
$\tilde{\Gamma}^i = - \partial_j \tilde{g}^{ij}$ can be used for the source terms 
without affecting well-posedness, but changing for example the source terms of
the constraint-propagation equations. 
In the BAM code we make the following choices:
 \begin{itemize} 

\item Wherever $\tilde{\Gamma}^i$ appears undifferentiated, we explicitly
use  $\tilde{\Gamma}^i = - \partial_j \tilde{g}^{ij}$ instead of the evolved
variable $\tilde{\Gamma}^i$. Otherwise, $\tilde{\Gamma}^i$ is used. 

\item The algebraic constraints $Tr(\tilde{A}_{ij}) = 0$ and
  $\det(\tilde{g}_{ij}) = 1$ are imposed whenever the right-hand sides are
  calculated, and also at the end of each evolution step. (Imposing the
  algebraic constraints at each evolution mini-step does not imply that they
  will hold after each full time step, because of the nonlinearity of the
  expressions involved.) 

\end{itemize}

Further important choices concern the treatment of the conformal factor and the
gauge choice. We will describe below the most popular choices that have been
used successfully in the literature,  and present some comparisons of different
choices in Sections ~\ref{sec:Single} and ~\ref{sec:Orbits}.

\subsection{Choices for the conformal factor}\label{Factor}

Let us first recall the fixed-puncture method, where the puncture
pole is treated analytically and the punctures do not move.
This is described in detail in \cite{Alcubierre02a}. 
The BSSN conformal factor is split according to
\beq
\label{fixedpuncture}
	\phi = \ln \psi = \xi + \ln{\psi_{BL}},
\eeq
where $\xi$ is assumed to be regular at the puncture. In an evolution
the regular function $\xi$ is evolved via the corresponding version of
Eq.~(\ref{phidot}), and the logarithmically singular part $\ln{\psi_{BL}}$ is 
kept constant. The key issue in the whole approach is to show that all
evolved variables remain sufficiently regular at the
punctures during evolution. In \cite{Alcubierre02a}, a detailed analysis is
given in terms of power counting arguments.

The disadvantage of this method is that it assumes a natural split of
the conformal factor according to Eq.~(\ref{fixedpuncture}) throughout an 
evolution, i.e., that the slices remain connected to all asymptotically
flat ends. 

In the moving-puncture approach the entire conformal factor is evolved. No
assumption is made about the geometry of the slices.  The slices are now
allowed to approach whatever geometry is preferred by the gauge conditions. It
turns out that in this preferred geometry the conformal factor does not
maintain the $1/r$ Brill-Lindquist pole, and instead develops a $1/\sqrt{r}$
pole at the ``puncture'' \cite{Hannam:2006vv}.
The puncture ceases to represent a second infinity, 
and instead corresponds to a surface inside the horizon.
The space outside this surface can be
accurately resolved with a finite-difference code. We can then regard the
moving-puncture method not as a mere trick to prolong the lifetime of a
black-hole evolution, but rather as an elegant and simple alternative to
excision techniques: the singularity is not cut out of the numerical grid, it
is avoided by the choice of gauge \cite{Hannam:2006vv}. 

The question now is how to evolve the divergent conformal factor. 
In practice two proposals have been found to work, which we will call the
$\phi$-method and the $\chi$-method. In the
$\phi$-method~\cite{Baker05a}, one works directly with the original
BSSN variable $\phi$,
\beq
	\phi = \ln{\psi},
\eeq
and the evolution system remains as Eqs~(\ref{phidot})-(\ref{Gdot}). The
purely experimental result is that finite differencing across the $\ln(r)$
singularity at $r=0$ leads to stable evolutions. 

In the $\chi$-method~\cite{Campanelli:2005dd}, a new conformal factor
is defined that is finite at the puncture,
\bea
	\chi &=& \psi^{-4},
\\	
\label{chidot}
	\partial_0 \chi &=& \frac{2}{3} \chi (\alpha K - \partial_j\beta^j).
\eea
Now Eq.~(\ref{chidot}) replaces Eq.~(\ref{phidot}) in the evolution system. 
If $\psi$ has the usual $1/r$ pole at the puncture, then $\chi =
O(r^4)$ at the puncture. As discussed in \cite{Hannam:2006vv},
the behavior changes to $\chi = O(r^2)$ during the evolution.

This approach does not rely on finite differencing of a singularity, but the
singular structure of the black hole is incorporated in the vanishing of
$\chi$ at the puncture. Because of divisions by $\chi$ present in the
evolution equations, care needs to be taken in the numerical implementation
to avoid divisions by zero or discontinuities arising out of unphysically
negative values of $\chi$.
We find that these problems can be avoided if $\chi$ is
consistently replaced in the right-hand sides of (\ref{phidot})-(\ref{Gdot})
by $\max(\chi,\epsilon)$ (for some small $\epsilon$) wherever divisions by
$\chi$ occur. As a general rule, we choose $\epsilon$ as follows. We know that
near the puncture, $\chi \sim (r/2m)^4$ in the initial data, and later evolves
to the form $\chi \sim (r/m)^2$. We therefore expect that $\chi$ will not fall
far below its initial minimum value, and choose $\epsilon$ to be less than
$(r_{min}/2m)^4$.

\subsection{Choices for the gauge}\label{Gauge}

The second ingredient in the moving-puncture method is a modification to the 
gauge choice. Both approaches now rely on the ``covariant'' form of ``1+log''
slicing \cite{Bona97a},
\beq
\label{oplwithshift}
	(\partial_t - \beta^i\partial_i) \alpha = -2\alpha K.
\eeq
The shift advection term had been dropped in the version of ``1+log'' slicing
used in the analytic fixed-puncture approach:
\beq
\label{opl}
	\partial_t \alpha = -2\alpha K, 
\eeq
and also in
the first version of the moving-puncture approach of \cite{Baker05a}. An attractive
feature of Eq.~(\ref{opl}) is that the slicing is asymptotically maximal for a
stationary solution, such as the final Kerr black hole of a merged
binary. However, Eq.~(\ref{opl}) admits undesirable zero-speed modes in the
BSSN system \cite{vanMeter:2006vi,Gundlach:2006tw} and
Eq.~(\ref{oplwithshift}) turns out to be a better choice for moving-puncture
evolutions \cite{Campanelli:2005dd,Baker:2006yw}. The stationary Schwarzschild
slicing with Eq.~(\ref{oplwithshift}) is given in
\cite{Buchman:2005ub,Hannam:2006vv}. 

For the shift, we use a gamma-freezing condition. The original gamma-freezing
condition introduced in \cite{Alcubierre02a} is 
\beq
\label{Gfreezing0}
	\partial_t \beta^i = \frac{3}{4} B^i, \quad
	\partial_t B^i  =  \partial_t \tilde \Gamma^i - \eta B^i.
\eeq Variants of this condition \cite{Baker05a,Baker:2006yw,vanMeter:2006vi}
consist of replacing some or all of the $\partial_t$ derivatives with
$\partial_0 = \partial_t - \beta^i \partial_i$. We will label these options
with reference to each of the three time derivatives in
(\ref{Gfreezing0}): ``ttt'' denotes that $\partial_t$ is used for all three
derivatives, ``0tt'' denotes that $\partial_0$ is used for the first time
derivative, and $\partial_t$ for the other two, and so on. The properties of
the different choices are studied in \cite{Gundlach:2006tw,vanMeter:2006vi}.
Reference \cite{Gundlach:2006tw}  proves that the combination of the BSSN equations 
with the ``1+log'' slicing condition (\ref{oplwithshift})
and the $000$ shift choice is strongly hyperbolic in the
sense of first order in time, second order in space systems
\cite{Nagy:2004td,Gundlach04a,Gundlach:2004jp, 
Gundlach:2005ta,Calabrese:2005ft}, 
and thus yields a well-posed initial-value problem. For the final results
presented in Section \ref{sec:Orbits} we quote results 
obtained with the ``ttt'' and ``000'' options, which are both found to yield
stable evolutions. All our recent work is based on the manifestly hyperbolic
choice ``000'', i.e., we make the replacement $\partial_t \rightarrow
\partial_0$ everywhere.

\section{Asymptotics}\label{sec:asymptotics}

\subsection{Using $\Psi_4$ for wave extraction}\label{sec:waves}

Extracting physical information from numerical simulations in general
relativity represents a highly non-trivial task for two reasons. First,
most of the functions numerically computed in the course of an evolution
are inherently coordinate dependent. Second, quantities commonly used
for the description of local systems in other areas of physics, such
as energy and angular momentum, are hard to define in an unambiguous
way in corresponding scenarios in general relativity. For the problem
at hand, the most important physical information to be extracted are
the energy and momenta radiated away in the form of gravitational
waves and the precise shape of these gravitational waves as seen
by a detector at large distances from the source.

In the past, the extraction of these quantities from numerical
simulations has been performed using either the Zerilli-Moncrief
(see e.\,g.\,\cite{Nagar05}) or the Newman-Penrose approach. In this
work we focus on the calculation of the Newman-Penrose scalar $\Psi_4$.
This method has been discussed frequently in the literature, but we provide a
detailed description to make clear the conventions we use (which can lead, for
example, to differences in signs or constant factors of two), and to provide a
complete, self-contained account of all the steps involved in calculating
waveforms as well as radiated momenta and energy from the numerically evolved
variables. For this purpose we will assume as known on a given hypersurface
$t=\mathrm{const}$ the ADM variables $g_{ij}$ and $K_{ij}$.

The Newman-Penrose scalar $\Psi_4$ is defined by
\begin{equation}
  \Psi_4 = -R_{\alpha \beta \gamma \delta} n^{\alpha} \bar{m}^{\beta}
      n^{\gamma} \bar{m}^{\delta}, \label{eq: psi4}
\end{equation}
where $R_{\alpha \beta \gamma \delta}$ represents the four-dimensional
Riemann tensor (with the sign convention of \cite{Misner73}) and $n$,
$\bar{m}$ form part of a null-tetrad $\ell$, $n$, $m$,
$\bar{m}$. Specifically, $\ell$ and $n$ denote ingoing and outgoing
null-vectors whereas the complex-valued $m$ is constructed out of two
spatial vectors orthogonal to $\ell$ and $n$, such that
\begin{equation}
  -\ell \cdot n =1 = m\cdot \bar{m},
\end{equation}
and all other inner products between the four-vectors vanish. 
$\Psi_4$ transforms as a
spin-weight $-2$ field (that is, under tetrad rotations which leave
$\ell$ and $n$ unchanged and rotate $m$ and $\bar{m}$ by an angle
$\theta$, we have $\Psi_4\rightarrow e^{-2i\theta} \Psi_4$).
Such objects represent symmetric trace-free
tensor fields on a sphere (in our case $R_{\alpha \beta \gamma \delta}
n^{\alpha} n^{\gamma}$) in terms of a complex scalar field.
For a quick introduction to spin-weighted
fields see e.g., \cite{Gomez97}.
There remains
freedom in the choice of  tetrad used in defining $\Psi_4$. Here, we first construct a
triad of orthonormal spatial vectors by applying the Gram-Schmidt
orthonormalization procedure to the three-dimensional vectors
\begin{eqnarray}
  u^i &=& \left[ -y,\,x,\, 0 \right], \nonumber \\
  v^i &=& \left[ x,\,y,\,z\right], \nonumber \\
  w^i &=& g^{ia} \epsilon_{abc}u^a v^b. \label{eq: triad}
\end{eqnarray}
The tetrad vectors are then given by
\begin{eqnarray}
  & n^0 = \frac{1}{\sqrt{2}\alpha} \qquad
     & n^{i} = \frac{1}{\sqrt{2}} \left(\frac{-\beta^i}{\alpha} - v^i\right),
       \label{eq: tetrad_n} \\
  & l^0 = \frac{1}{\sqrt{2}\alpha}
     & l^{i} = \frac{1}{\sqrt{2}} \left(\frac{-\beta^i}{\alpha} + v^i\right),
       \label{eq: tetrad_l} \\
  & m^0 = 0
     & m^i = \frac{1}{\sqrt{2}} \left( u^i + iw^i \right).
       \label{eq: tetrad_m}
\end{eqnarray}
Next, we need to express Eq.\ (\ref{eq: psi4}) in terms of the
three-dimensional quantities available on each time slice. This is achieved
by virtue of the Gauss-Codazzi and the Mainardi equations which relate
the space-time projections of the four-dimensional Riemann tensor to
its three-dimensional counterpart and the ADM variables according to
\begin{eqnarray}
  \bot R_{\alpha \beta \gamma \delta} &=& \mathcal{R}_{\alpha \beta \gamma
       \delta} + K_{\alpha \gamma} K_{\beta \delta} - K_{\alpha \delta}
       K_{\beta \gamma}, \\
  \bot R_{\mu \beta \gamma \delta} \hat{n}^{\mu} &=&
       D_{\gamma} K_{\beta \delta} - D_{\delta}K_{\beta \gamma}, \\
  \bot R_{\mu \beta \nu \delta} \hat{n}^{\mu} \hat{n}^{\nu} &=&
       \mathcal{R}_{\beta \delta} - K_{\beta \mu}K^{\mu}{}_{\delta}
       + K\, K_{\beta \delta},
\end{eqnarray}
where $\mathcal{R}_{\alpha \beta \gamma \delta}$ is the three-dimensional
Riemann tensor,
$\hat{n}^{\alpha}$ the timelike unit normal vector associated with the
foliation and
we follow York's \cite{York79} notation for the projection operator $\bot$,
which is, for example for an arbitrary tensor $T_{\alpha \beta}$,
\begin{equation}
  \bot T_{\alpha \beta}:=
       (\delta^{\mu}{}_{\alpha}+\hat{n}^{\mu}\hat{n}_{\alpha})
       (\delta^{\nu}{}_{\beta}+\hat{n}^{\nu}\hat{n}_{\beta}) T_{\mu \nu}.
       \nonumber
\end{equation}
In our coordinate basis adapted to the ``3+1'' decomposition, we
are thus able to express $\Psi_4$ exclusively in terms of the
ADM variables as well as the triad vectors constructed from (\ref{eq: triad})
according to
\begin{eqnarray}
  \Psi_4 &=& -\frac{1}{4} \left( R_{abcd} v^a v^c
           - 2 \bot R_{\alpha bcd} \hat{n}^{\alpha} v^c
           + \bot R_{\alpha b \gamma d} \hat{n}^{\alpha} \hat{n}^{\gamma}
           \right) \nonumber \\
         && (u^b-iw^b)(u^d-iw^d).
\end{eqnarray}
The contributions of the individual modes $\ell$, $m$
are obtained from projecting
$\Psi_4$ onto the spherical harmonics $Y^{-2}_{\ell m}$
of spin weight $-2$. These projections
are defined in terms of the scalar product
\begin{equation}
  A_{\ell m} = \langle Y^{-2}_{\ell m}, \Psi_4 \rangle =
      \int_0^{2\pi} \int_0^{\pi}
      \Psi_4 \overline{Y^{-2}_{\ell m}}\, \sin \theta d\theta d\phi
      \label{eq: scalar_product}
\end{equation}
which, in practice, is evaluated at a finite extraction radius
$r_{ext}$. 

The spin-weighted spherical harmonics $Y^{s}_{\ell m}$ can be defined
in terms of the Wigner $d$-functions (e.g.\ \cite{Wiaux:2005fm}) as
\begin{equation}
Y^{s}_{{\ell}m}\left(\theta,\varphi\right)
=
\left(-1\right)^{s}\sqrt{\frac{2{\ell}+1}{4\pi}}d_{m(-s)}^{{\ell}}\left
(\theta\right)e^{im\varphi},
\end{equation}
where
\begin{eqnarray}
&& d_{ms}^{{\ell}}\left(\theta\right) 
\nonumber \\
& = & 
\sum_{t=C_{1}}^{C_{2}}\frac{\left(-1\right)^{t}\left[\left({\ell}+m\right)!
\left({\ell}-m\right)!\left({\ell}+s\right)!\left({\ell}-s\right)!\right]^{1/2}}
{\left({\ell}+m-t\right)!\left({\ell}-s-t\right)!t!\left(t+s-m\right)!}
\nonumber \\
 &  &
\left(\cos\theta/2\right)^{2{\ell}+m-s-2t}\left(\sin\theta/2\right)^{2t+s-m},
\end{eqnarray}
with $C_{1}=\max(0,m-s)$ and $C_{2}=\min({\ell}+m,{\ell}-s)$.
For $\ell=2$ and spin-weight $s=-2$, we have
\begin{eqnarray}
  Y^{-2}_{2-2} &=& \sqrt{\frac{5}{64\pi}} \left( 1 -\cos \theta \right)^2
       e^{-2i\phi}, \nonumber \\
  Y^{-2}_{2-1} &=& -\sqrt{\frac{5}{16\pi}} \sin \theta \left(
       1-\cos \theta \right) e^{-i\phi}, \nonumber \\
  Y^{-2}_{20} &=& \sqrt{\frac{15}{32\pi}} \sin^2 \theta, \nonumber \\
  Y^{-2}_{21} &=& -\sqrt{\frac{5}{16\pi}} \sin \theta \left(
       1+\cos \theta \right) e^{i\phi}, \nonumber \\
  Y^{-2}_{22} &=& \sqrt{\frac{5}{64\pi}} \left( 1 +\cos \theta \right)^2
       e^{2i\phi}.
\end{eqnarray}
In practice, the integrand on the right-hand side of Eq.\ (\ref{eq:
scalar_product}) is evaluated on the Cartesian grid and interpolated
onto a sphere of extraction radius $r_{ext}$ using fifth-order
polynomials. The integration over the sphere is performed using the
fourth-order Simpson method.

While this procedure is straightforward
from a numerical point of view, we emphasize one delicate point. In order
to reduce the computational costs, numerical simulations are
often performed with explicit use of symmetry properties of the
spacetime under consideration. For this purpose it is important to
take into account the transformation of the variables under inversion
of the $x$, $y$ or $z$ coordinate. In our case the non-trivial operation
is the symmetry across the $xy$ plane. This problem manifests itself in
the calculation of the modes according to (\ref{eq: scalar_product}) where
the integrand is directly available only in the range $0\le \theta \le \pi/2$.
Using the parity properties of the functions involved, however, we are
able to transform the right hand side of Eq.\ (\ref{eq: scalar_product})
into an integral restricted to the northern hemisphere.
In particular, in the case of reflection in the $z$-direction
($(x,y,z)\rightarrow (x,y,-z)$), which is relevant here,
the real part of $\Psi_4$ behaves like an even function, whereas the
imaginary part of $\Psi_4$ behaves as an odd function.

Similarly, the harmonics $Y^{-2}_{22}$, $Y^{-2}_{2-2}$ transform into
the complex conjugates of each other. In summary,
\begin{eqnarray}
  \Psi_4(\pi -\theta,\phi) &=&  \overline{\Psi_4(\theta,\phi)} \\
  Y^{-2}_{22}(\pi -\theta,\phi) &=& \overline{Y^{-2}_{2-2}(\theta,\phi)} \\
  Y^{-2}_{2-2}(\pi -\theta,\phi) &=& \overline{Y^{-2}_{22}(\theta,\phi)}
\end{eqnarray}
We use the following relation valid for arbitrary functions of $\theta$
\begin{equation}
  \int_0^{\pi} f(\theta) d\theta = \int_0^{\pi/2} \left[f(\theta) + f(\pi -\theta)\right] d\theta,
\end{equation}
and are thus able to calculate
\begin{eqnarray}
  A_{22} &=& \langle Y^{-2}_{22}, \Psi_4 \rangle =
      \int_0^{2\pi} \int_0^{\pi}
      \Psi_4 \overline{Y^{-2}_{22}}\, \sin \theta d\theta d\phi
      \\
    &=& \int_0^{2\pi} \int_0^{\pi/2} \left( \Psi_4
        \overline{Y^{-2}_{22}} + \overline{\Psi_4}
        Y^{-2}_{2-2} \right) \sin \theta d\theta d\phi \nonumber
\end{eqnarray}
An equivalent change of basis to represent functions on the sphere
has been discussed by Zlochower {\em et al.}~in \cite{Zlochower03}.

In the study of numerical simulations of black-hole-binary systems, one
is often interested in the amount of energy and momenta radiated away
from the system in the form of gravitational radiation. In terms of the
Newman-Penrose scalar $\Psi_4$, these are given by the expressions
\begin{eqnarray}
  \frac{dE}{dt} &=& \lim_{r\rightarrow \infty} \left[ \frac{r^2}{16\pi}
      \int_{\Omega} \left| \int_{-\infty}^{t} \Psi_4 d\tilde{t}
      \right|^2 d{\Omega}\right], \\
  \frac{dP_i}{dt} &=& -\lim_{r \rightarrow \infty} \left[ \frac{r^2}{16\pi}
      \int_{\Omega} \ell_i \left| \int_{-\infty}^{t} \Psi_4 d\tilde{t}
      \right|^2 d{\Omega}\right], \\
  \frac{dJ_z}{dt} &=& -\lim_{r \rightarrow \infty} \left\{ \frac{r^2}{16\pi}
      \mathrm{Re}\left[ \int_{\Omega} \left( \partial_{\phi} \int_{-\infty}^t
      \Psi_4 d\tilde{t}
      \right)
      \right. \right. \nonumber \\
      &&\left. \left.
      \left(
      \int_{-\infty}^t \int_{-\infty}^{\hat{t}}
      \overline{\Psi_4} d\tilde{t} d\hat{t}
      \right)d{\Omega} \right] \right\},
\end{eqnarray}
where
\begin{equation}
  \ell_i = \left(-\sin \theta \cos \phi,\,\,\,-\sin \theta \sin \phi,\,\,\,
           -\cos \phi \right).
\end{equation}
We have listed these relations explicitly, because of different conventions
in use in the literature. In particular we emphasize the difference
by a factor of $1/4$ with Eqs.\ (22)-(24) of \cite{Campanelli99} which
arises out of differences in the scaling of the tetrad vectors
[cf.\ our Eqs.\ (\ref{eq: tetrad_n})-(\ref{eq: tetrad_l}) with their
Eq.\ (30)].

The expression for the energy can be simplified by using the expansion
of $\Psi_4$ in modes $\ell$, $m$. Taking into account the
orthonormality of the spin-weighted harmonics we obtain
\begin{equation}\label{eq:Edot}
  \frac{dE}{dt} = \lim_{r \rightarrow \infty} \left[ \frac{r^2}{16\pi} \left|
      \int_{-\infty}^t \sum_{l,m} A_{\ell m} d\tilde{t} \right|^2 \right].
\end{equation}
In particular, this relation enables us to calculate the energy radiated
in an individual mode. For the equal-mass systems considered in this work,
we find $>99\,\%$ of the energy to be radiated in the form of the dominant
$\ell=2$, $m=\pm 2$ modes.

Finally, let us note that in analyzing waveform modes as functions of time, it
is extremely useful to split the complex function representing $r \Psi_4$ (or,
say, the strain $h$) as
\begin{equation}
\label{eq:phase_angle}
r \Psi_4 (t)  = A(t) e^{i \varphi(t) }, \quad \omega(t) = \frac{\partial
  \varphi(t)}{\partial_t}, 
\end{equation}
as suggested in \cite{Baker:2006yw}.
In this paper this representation proves particularly useful to compare
different initial data sets as in \ref{sec:parameters}.

\subsection{Total energy, linear and angular momentum}\label{sec:ADM}

In general relativity unambiguous notions 
of the energy-momentum four-vector and angular momentum can only be assigned
to a spacetime as global quantities, determined from the asymptotic structure
of the spacetime. In this sense two types of quantity can be defined: those
that are conserved by time evolution, and those that decrease with time,
expressing the radiation of energy-momentum and angular momentum to infinity. 

The expression for the energy-momentum at spatial infinity, which is
time-independent and which corresponds to a four-vector under Lorenz
transformations, was given first by Arnowitt, Deser and Misner in 1962
\cite{Arnowitt62} in the context of the Hamiltonian formalism. This quantity
is usually called the ADM energy-momentum, the time component being called the
ADM energy or, somewhat inconsistently, the ADM mass, different from the rest
mass to be defined below.
The expressions can be given as limits of surface integrals defined at finite
radius, and are evaluated in asymptotically Cartesian (regular)
coordinates $\{x^{i}\}$ --- where the components of the spatial metric tend to 
diag$(1,1,1)$ for large radii. The surfaces are then taken as spheres $S_{r}$
of radius $r$. 

We define the surface integrals (which we will also refer to as ADM integrals)
\begin{eqnarray}
E(r) & = & \frac{1}{16\pi}
\int_{S_{r}}\sqrt{g}g^{ij}g^{kl}\left(g_{ik,j}-g_{ij,k}\right)dS_{l},
\label{madm_int} \\ 
P_{j}(r) & = & \frac{1}{8\pi}\int_{S_{r}}\sqrt{g}\left (K^{i}_{j} -
  \delta^{i}_{j}K \right)dS_{i}, \label{padm_int} \\ 
J_{j}(r) & = & \frac{1}{8\pi} \epsilon_{jl}{}^m \int_{S_{r}} \sqrt{g} x^l
\left(K^i_m - K \delta^i_m \right) dS_{i} \label{jadm_int} 
\end{eqnarray}
which  have to be evaluated in an asymptotically Cartesian coordinate system.

The ADM energy $M_{ADM}$ and linear and angular momentum $P_j$ and $J_j$ are
then given by \cite{Omurchadha74,York:1974hp,York79}
\begin{eqnarray}\label{madm}
M_{ADM} & = & \lim_{r\rightarrow\infty} E(r),  \\
P_{j} & = & \lim_{r\rightarrow\infty} P_{j}(r), \\
J_{j} & = & \lim_{r\rightarrow\infty} J_{j}(r)
\end{eqnarray}
and the rest mass $M_{R}$ can be defined as $M_{R}^2=M_{ADM}^{2}-\sum_{j=1,3}P_{j}P_{j}$.

For radiation processes we also require definitions of total energy, linear
and angular momentum that decrease as energy and linear as well as angular
momentum are radiated to infinity. The appropriate quantities are the Bondi
quantities \cite{Bondi62}, which can be defined as taking the limit of the ADM
integrals not toward spatial, but rather null, infinity
\cite{Katz88,Brown:1996bw,Poisson04a}, i.e., the limit to infinite distance is
taken for constant retarded time instead of on a fixed Cauchy slice. In the
context of our numerical treatment, the ADM and Bondi quantities can be computed rather
accurately by computing values at several radii, and then performing a
Richardson extrapolation (in extraction radius, not, as is more usual, in grid spacing) 
to infinity. Here the Bondi quantities can be computed at
any time for a fixed extraction radius, and have to be compared between
different radii by taking into account the light travel time between the
timelike cylinders of different radii. This time delay can be estimated 
from a corresponding Schwarzschild solution as is done in \cite{Fiske05}
by the difference in the values of the radial ``tortoise coordinate'' values as
\begin{equation}
\Delta T(R_1, R_2) = \left[ R + 2 M \ln (R/2 M - 1)\right]_{R=R_1}^{R=R_2},
\label{eq:Tortoise} 
\end{equation}
where the radii $R_i$ are understood as Schwarzschild radius (i.e.\ luminosity distance),
and the Schwarzschild radius can be estimated from the simulation's radial
coordinate $r$ by assuming it corresponds to  the isotropic radial 
coordinate in Schwarzschild spacetime, which yields  $R = r (1 + 2M/r)^2$.

\section{Numerical Method}\label{sec:Numerics}

The numerical method of our black-hole simulations is based on a
method of lines approach using finite differencing in space and
explicit Runge-Kutta (RK) time stepping.  For efficiency,
Berger-Oliger type adaptive mesh refinement (AMR) is
used~\cite{Berger84}. The new numerical results discussed in this
paper were obtained with the BAM
code~\cite{Bruegmann96,Bruegmann97,Bruegmann:2003aw}, which implements
a particular AMR strategy that we describe below (we also compare with
published results obtained with Sperhake's LEAN
code~\cite{Sperhake:2006cy}). Although BAM also includes an
experimental oct-tree cell based algorithm that allows arbitrarily
shaped refinement levels, this has not been used since a simpler box
based algorithm is sufficient for black-holes binaries.

The numerical domain is represented by a hierarchy of nested Cartesian
grids. The hierarchy consists of $L$ levels of refinement
indexed by $l = 0, \ldots, L-1$. A refinement level consists
of one or more Cartesian grids with constant grid-spacing $h_l$ on
level $l$. A refinement factor of two is used such
that $h_l = h_0/2^l$. The grids are properly nested in that the
coordinate extent of any grid at level $l$, $l>0$, is completely
covered by the grids at level $l-1$. Of special interest are the
resolutions $h_{max}=h_0$ of the coarsest, outermost level, 
and $h_{min} = h_{L-1}$ of the finest level. 

Since we focus on the case of one or two black holes, a particularly
simple grid structure is possible where each refinement level consists
of exactly one or two non-overlapping grids. While the size of these
grids could be determined by truncation error estimates or some field
variable that indicates the need for refinement, for the purpose of
convergence studies we have found it convenient to specify the size of
the grids in advance. This allows, for example, the doubling of
resolution within a predetermined coordinate range. Concretely, let
$N_l$ be the number of points in any one direction for a cubical box
with $N_l^3$ points on level $l$. On level $l$, center such a box on
each of the black-hole punctures. If there are two punctures and the
two boxes do not overlap, this is the layout that is used. If two
boxes overlap, replace them by their bounding box, which is the
smallest rectangular (in general non-cubical) box that contains the
two original boxes. 

Assuming $N_l = N$ (a constant independent of $l$), a typical
configuration around two punctures consists of two separate cubical
boxes at $l=L-1$, but for decreasing $l$ and increasing $h_l$ the size
of the boxes increases until starting at some intermediate level the
boxes overlap and a single rectangular box is formed, which towards
$l=0$ becomes more and more cubical. 

The hierarchy of boxes evolves as the punctures move. We use the shift
to track the position $x^i_{punc}$ of a puncture by integrating
\beq
	\partial_t x^i_{punc} = - \beta^i(x^j_{punc}),
\eeq
cf.~\cite{Campanelli:2005dd}, using the ICN method.
The outermost box on level 0 and also several of the next finer levels are
chosen to be single cubes of fixed size centered on the origin to avoid
unnecessary grid motion. 

Note that as long as one neglects the propagation of gravitational waves,
the nesting described above represents in a natural manner
the $1/r$ fall-off of the metric for a single puncture. For a single
puncture and fixed $N$, doubling the grid-spacing going from level $l$
to $l-1$, i.e., $h_{l-1}=2h_l$, puts the boundary of a centered cube
twice as far away. If a resolution of $h_l$ is sufficient to
resolve the metric at $1/r$, then $2h_l$ should be sufficient to
resolve the metric at $1/(2r)$ since this is the slowest fall-off of
any metric variable.
This was the rationale for the nested box fixed
mesh refinement (FMR) introduced in~\cite{Bruegmann96}, which was
found to work well in practice for the first 3d mesh-refinement
evolutions of black holes~\cite{Bruegmann97}. This FMR nesting
strategy generalizes straightforwardly to the case of two moving
punctures as outlined above.

In the presence of gravitational waves further demands for spatial
resolution arise: the wave amplitude falls off with $1/r$,
corresponding to the roughly constant amplitude of the ``predicted''
signal $r \psi_4$, while the wavelength is approximately constant. The
spatial profile of the signal thus requires constant radial resolution
with increasing distance, while the amplitude fall-off leads to
increasing accuracy requirements as distance increases, in order to
separate the waves from the background. Correspondingly, the grids
need to be adapted when waves need to be traced accurately to typical
wave extraction distances, which in a setup as presented here are
still rather limited by computational cost. In actual runs it is thus
convenient to use at least two different values for the $N_l$, one for
the cubes that resolve the neighborhood of the punctures, another one
for the levels where the wave extraction is performed. For
Berger-Oliger time-stepping most of the computational work is
performed on the finest levels, so one chooses $N_{L-1}$ as small as
possible (while still covering the entire black hole with sufficiently
fine resolution), and we can gain some extra resolution for wave
extraction at small extra cost by using a larger box for the levels on
which waves are extracted.

The grids are cell-centered. For example, in one dimension for the
cell given by the interval $[0,h_0]$, the data on level 0 is located
at the point $h_0/2$, on level 1 at $h_0/4$ and $3h_0/4$, on level 2
at $h_0/8$, $3h_0/8$, $5h_0/8$, and $7h_0/8$, and so forth.
Data is transferred between levels by sixth-order polynomial interpolation,
where the three-dimensional interpolant is obtained by successive
one-dimensional interpolations.

On any given box with resolution $h_l$, we implement fourth-order
finite differencing for the spatial derivatives of the Einstein
equations. Standard centered stencils are used for all first and second-order
derivatives except for advection derivatives, $\beta^i
\partial_i$. For second-order finite differencing, the advection terms
required one-sided differencing for stability. For fourth-order finite
differencing, we found that both centered and one-sided differencing
can lead to severe instabilities with ICN time stepping, 
while ``lop-sided'' stencils
lead to stable evolutions (cf.~\cite{Zlochower2005:fourth-order}). Our
runs are performed using such lop-sided advection derivatives with
fourth-order Runge-Kutta (RK4).

The code allows us to add artificial dissipation terms to the right-hand-sides
of the time evolution equations, schematically written as
\begin{equation}
\partial_t {\bf u} \rightarrow \partial_t {\bf u} + Q {\bf u},
\end{equation}
in particular we use the standard Kreiss-Oliger dissipation
\cite{Kreiss73,Gustafsson95} 
operator ($Q$) of order $2r$
\begin{equation}
Q = \sigma (-h)^{2 r - 1} (D_+)^{r} \rho \, (D_-)^{r}/2^{2r},
\end{equation}
for a $(2r -2)$-accurate scheme, with $\sigma$ a parameter regulating
the strength of the dissipation, and $\rho$ a weight function that we
currently set to unity. Adding artifical dissipation is apparently not
required for stability in our runs, but we have used dissipation for
RK4 evolutions to avoid high frequency noise from mesh-refinement
boundaries.  We find that the inherently stronger dissipation of the
ICN algorithm also rather efficiently suppresses noise from refinement
boundaries, and our ICN test runs suggest that in this case the adding
of Kreiss-Oliger dissipation is superfluous.

All AMR results for two punctures reported so far are based
on codes that involve at least some second-order component, while BAM
in principle allows fully fourth-order AMR. In particular, we apply
sixth order polynomial interpolation in space between different
refinement levels so that all spatial operations of the AMR method are
at least fourth order. However, there are three sources of second-order
errors. One is the initial data solver, although this initial
error appears to be negligible in the cases we consider here.  Another
source of second-order error is the implementation of the radiative
boundary condition. However, the nested boxes position the outer
boundary at sufficiently large distances such that these errors do not
contribute significantly (ideally because they are causally
disconnected from the wave extraction zone). The final source of
second-order error in our current runs is due to interpolation in time
within the Berger-Oliger time-stepping scheme, which is worth
discussing in some more detail.

Berger-Oliger time-stepping can be stated as recursive
pseudo-code for example as:
\begin{quote}
evolve\_hierarchy($l$, $\Delta t$)
\\
\hspace*{5mm} evolve($l$, $\Delta t$)
\\
\hspace*{5mm} if ($l+1<L$) 
\\
\hspace*{10mm} evolve\_hierarchy($l+1$, $\Delta t/2$)
\\
\hspace*{10mm} evolve\_hierarchy($l+1$, $\Delta t/2$)
\\
\hspace*{5mm} if ($l>0$)
\\
\hspace*{10mm} restrict\_prolong($l-1$, $l$)
\\
\hspace*{5mm} regrid($l$)

\end{quote}
The recursion is started by calling the function ``evolve\_\-hierarchy'' for
$l=0$, i.e., beginning with the coarsest level.
The function ``evolve\_hierarchy'' evolves all levels from $l$ to
$L-1$, the finest level, by a time step of $\Delta t$ forward in time. 
First, level $l$ is evolved by $\Delta t$ by the function ``evolve''.
Then the function ``evolve\_hierarchy'' calls itself recursively to
advance level $l+1$ and all its sublevels twice by $\Delta t/2$.
The recursion ends if level $l+1$ does not exist, i.e., if $l+1$
is not less than $L$, then ``evolve\_hierarchy'' does not call itself
again. 
Once all levels $l$ through $L-1$ have reached the next time level,
information is exchanged between levels $l-1$ and $l$, denoted by a
call to ``restrict\_prolong'' if $l>0$. In particular, the refinement
boundary of $l$ is populated using information from $l-1$. The result
is the new level $l$. Finally, the refinement hierarchy is updated by
the function ``regrid''.

Although the time stepping used for evolution is fourth-order
Runge-Kutta, there arises the additional issue of how to provide
boundary values for the intermediate time-levels of the Berger-Oliger
algorithm that are not aligned in time with a coarser level
(otherwise spatial interpolation can be used). There are several
options for fourth-order boundaries.

The original suggestion by Berger and Oliger is to interpolate in time
(over several coarse levels at different instances of time) in order
to obtain boundary values for a fine level. One can use three time levels
of the coarser level to perform quadratic interpolation (third order
in the time step) resulting in overall second-order convergence when
using a leapfrog scheme, e.g., as done in~\cite{Bruegmann96}. However,
the convergence order and the stability of the algorithm depends on
the form chosen for the Einstein equations and on the time-stepping
algorithm used. For example, quadratic interpolation for ICN and a
first order in time, second order in space formulation can lead to a
drop of convergence order and instabilities, see Schnetter et
al.~\cite{Schnetter-etal-03b}. Other authors report success with
different variants of time interpolation,
e.g.,~\cite{Pretorius:2005amr,Baker:2005xe}.

An alternative approach is to replace the single point refinement
boundary by a buffer zone consisting of several points,
e.g.,~\cite{Schnetter-etal-03b,Lehner:2005vc,Csizmadia:2006rz}.  The buffer
zone approach can be expected to perform well for the transmission of
waves through refinement boundaries, see e.g.,~\cite{Lehner:2005vc}
(note that special methods like~\cite{Baker:2005xe} seem to achieve similar
performance). The optimal number of buffer points is method dependent.
For example, RK4 requires 4 source evaluations, and if the lop-sided
stencil with 3 points in one direction is used, then the numerical
domain of dependence for a given point has a radius of 12
points. Therefore, it is possible to provide 12 buffer points at the
refinement boundary and to perform one RK4 time step with size 3
stencils that does not require any boundary updates. Only after the
time step is completed, the buffer zones have to be repopulated. In
the context of Berger-Oliger AMR, the buffer update is based on
interpolation from the coarser levels. Since every second time step at
level $l$ coincides in time with level $l-1$, one can provide 24
buffer points, perform two time steps, and then update the buffer by
interpolation in space. With 12 buffer points, one can interpolate in
time to obtain data for the buffer points at intermediate time levels.

For the simulations reported here, our standard setup is to use RK4
with dissipation and lop-sided advection stencils, 6 buffer points,
quadratic interpolation in time, and Berger-Oliger time-stepping on
all but the outermost grids.  Let us comment on these choices.

For some grid configurations we have encountered instabilities for
very large, coarse grids, that experimentally are connected to the
large time steps on the coarse grid. We were able to cure these
instabilities by turning-off Berger-Oliger time-stepping for the
outermost grids (cf.~\cite{Bruegmann:2003aw} where this idea was
introduced in a different context).

To use fewer than 12 buffer points, we can interpolate into all buffer
points before starting a RK4 update as described, and then evolve all
points except the outermost points located exactly on the boundary,
which are kept fixed at their initial interpolated value. The inner
points next to the boundary are updated using second order finite
differencing for the centered derivatives and shifted advection
stencils for the advection derivatives.
Experimentally, using just 6 buffer points leads to very small
differences compared to 12 buffer points, however smaller buffer zones
lead to noticeable differences.
Even though for large grids the number of buffer zones becomes
negligible, for the grid sizes that we have to use, the buffer points
impact the size of the grids significantly. For example, for a box of
size 64 in one direction, adding 6 points on both sides instead of 12
points corresponds to a savings of 35\% in the total number of
points. For clarity, we always quote grid sizes without buffer points,
because this is the number of points owned by a particular grid.

Using quadratic interpolation in time is, apart from the outer
boundary treatment, the only source of second-order errors in the
evolution scheme. We checked for a few cases that running without
Berger-Oliger time-stepping entirely led to only small differences
compared to other error sources.  However, for sufficiently high
resolutions, quadratic interpolation in time should become the
dominant error.  In principle, we can resolve this issue by either not
using Berger-Oliger time-stepping or by using larger buffer zones,
which at the moment is prohibitively expensive in resources.

We have also experimented with higher order in time interpolation,
although a systematic analytical and numerical analysis beyond these
first experiments is needed.  Simply using additional coarse time
levels was not successful. In general, if at time $t$ a fine level $l$
is not aligned in time with the coarser level $l-1$, we use the grid
functions on level $l-1$ at different times to interpolate to time
$t$. For quadratic interpolation these different times are $t+\Delta
t$, $t-\Delta t$, and $t-3\Delta t$, where $2\Delta t$ is the timestep
on level $l-1$. As mentioned before this kind of interpolation leads
to overall second-order accuracy in time at the interpolated
points. We routinely use this approach and it leads to stable
evolutions. In order to obtain a fully fourth-order scheme we have
included additional coarse levels at times $t-5\Delta t$ and
$t-7\Delta t$. However, this extended interpolation scheme over five
different times leads to oscillations at the refinement boundaries,
which are the points where we use interpolation in time. These
oscillations increase with resolution and are thus likely
instabilities which would cause the code to fail at sufficiently high
resolution. At the resolution considered in this paper these
instabilities do not cause the code to fail. However, they are a
significant source of noise, which propagates out of the refinement
boundaries into the rest of the grid. Since this noise is not
convergent, it eventually spoils convergence in the entire grid. One
reason for this problem may be the high degree of asymmetry in the
interpolation stencil which uses four points before time $t$ and only
one after $t$.

Finally, we note that BAM is MPI parallelized. The dynamic grid
hierarchy with moving and varying boxes introduces an additional
communication overhead compared to the FMR runs that BAM was used for
previously~\cite{Bruegmann:2003aw}. For up to 128 processors scaling seems
reasonable for a constant problem size per processor, but we do expect
issues for larger processor numbers, which we have not been able to
test yet.

\section{Single puncture with dynamic conformal factor}\label{sec:Single}

\subsection{Numerical experiments for a single stationary puncture}
\label{SingleNum}

In this section we apply the $\phi$ and $\chi$ moving-puncture methods to
evolutions of a single Schwarzschild puncture. This provides an excellent test
case, because we can compare with the analytic results in \cite{Hannam:2006vv},
and study the convergence properties of the code without the added
complication of moving mesh-refinement boxes. 

The initial data are as described in Section \ref{sec:ID}, where
$u = 0$ and $\tilde{A}_{ij} = 0$ on the initial 
slice, and we choose $m_1=M=1$. We use a "pre-collapsed" lapse of 
$\alpha = \psi^{-2}_0$ for these runs, but stress that similar
convergence properties are found with an initial lapse of $\alpha =
1$. The convergence series consist of evolutions with $N = 64^3, 96^3,
128^3$ grid points in each box, and seven levels of refinement below
the coarsest level (making a total of eight levels). The resolutions
on the coarsest levels are $h_{max} = 6M, 4M, 3M$, and the resolutions
on the finest levels and at the puncture are $h_{min} = 6M/128,
4M/128, 3M/128$. The gauge choice is $ttt$, and $\eta = 2.0/M$. For
these runs (and only for these) a uniform time-step was used on all
levels (i.e., not Berger-Oliger) in order to fully test the
fourth-order accuracy of the code.

\begin{figure}[t]
\centerline{\resizebox{9cm}{!}{\includegraphics{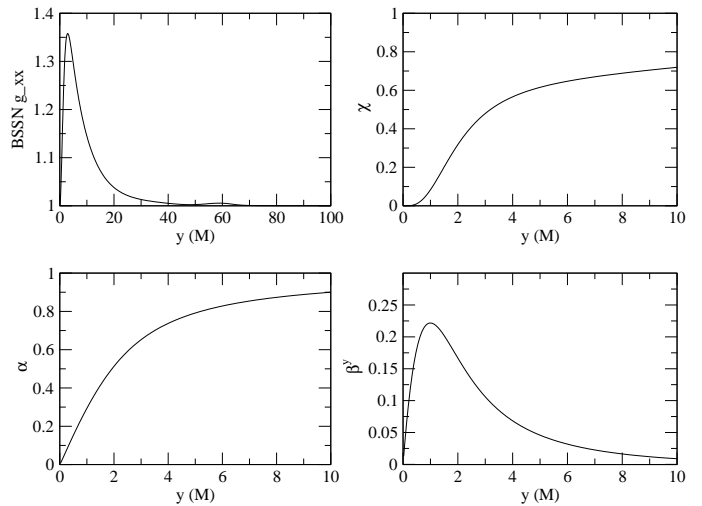}}}
\caption{The BSSN variables $\tilde{g}_{xx}$, $\chi$, $\alpha$, and $\beta^y$
  after $50M$ of evolution of a Schwarzschild puncture using the $\chi$
  method. A small pulse due to the initial adjustment of the gauge can be 
seen at about $y = 60M$ in $\tilde{g}_{xx}$. The main features of the other
variables are confined to $y < 10M$.}
\label{fig:chi_soln}
\end{figure}

As discussed in \cite{Hannam:2006vv}, a 1+log evolution of Schwarzschild reaches
a stationary slice, and at the puncture $\beta^2 = \beta_i\beta^i = 0.5239$ and the
Schwarzschild radial coordinate is $R = 1.3124M$. After $50M$ of evolution,
these values are reached to within $1.3\%$ and $0.5\%$ in the highest
resolution runs using the $\phi$ method. With the $\chi$ method, the errors in
$\beta^2$ and $R$ are $0.6\%$ and $0.2\%$. Figure \ref{fig:chi_soln} shows
several of the BSSN variables after $50M$ of evolution with the $\chi$ method. 

\begin{figure}[t]
\centerline{\resizebox{8.5cm}{!}{\includegraphics{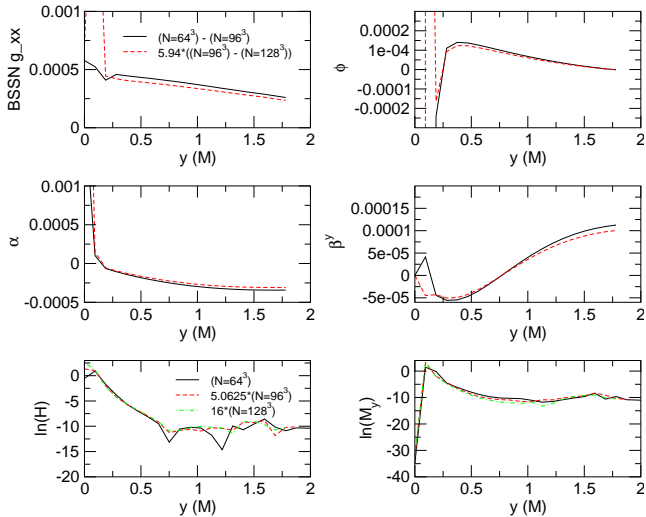}}}
\caption{Fourth-order convergence of a Schwarzschild puncture after $50M$ of
evolution, using the $\phi$ method. Results were taken from runs with $N = 64^3, 96^3$ and
$128^3$ points with octant symmetry. The plots show the differences between the
three runs, scaled to be consistent with fourth-order accuracy. For the 
Hamiltonian constraint and $y$-component of the momentum constraint, which
should converge to zero, we show the logarithm of the scaled values.}
\label{fig:cvg_phi_t50}
\end{figure}

The convergence of the $\phi$ method is demonstrated in Figure
\ref{fig:cvg_phi_t50}, which  shows convergence plots of $\tilde{g}_{xx}$,
$\phi$, $\alpha$,  $\beta^y$, and the Hamiltonian and 
momentum constraints.  The data are taken along the $y$-axis at $t = 50M$ on
the finest level of the mesh-refinement scheme. The errors in the Hamiltonian and  
momentum constraints cover a wide range, so the logarithm of the scaled errors
is shown. The differences are scaled assuming fourth-order convergence, and
the code demonstrates good fourth-order convergence everywhere except at the
points closest to the puncture. 

The $\phi$ variable shows extremely poor convergence at the puncture, but this
is to be expected: $\phi$ diverges like $\ln(r)$ near the puncture. What is
remarkable is that this non-convergent behavior remains localized at the
puncture, and does not affect the accuracy or stability of the evolution as a
whole. 

\begin{figure}[t]
\centerline{\resizebox{9cm}{!}{\includegraphics{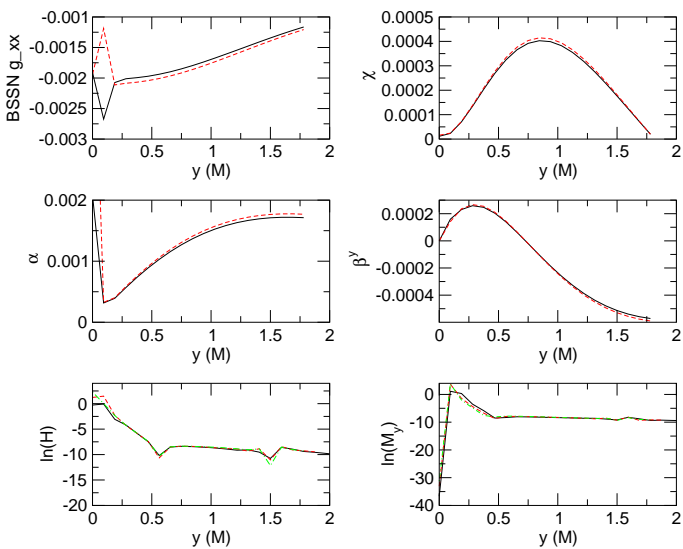}}}
\caption{Fourth-order convergence of a Schwarzschild puncture after $50M$ of
evolution, using the $\chi$ method. The parameters match those used for the runs
discussed in Figure \ref{fig:cvg_phi_t50}.}
\label{fig:cvg_chi_t50}
\end{figure}

Figure \ref{fig:cvg_chi_t50} shows similar convergence plots for the $\chi$
method. In this case the $\chi$ variable, which should behave like $r^2$ near the
puncture, is seen to converge everywhere. The constraints and $\beta^y$ also
show better convergence properties near the puncture. This is consistent with
the comparison with the stationary 1+log solution, where we see that the
$\chi$ method was more accurate at the puncture. 

We draw three conclusions from these results. (1) Our code is fourth-order
accurate for the resolutions used in this work,
at least when the mesh-refinement boxes do not move and a uniform
time-step is used. (2) The moving-puncture method extremely accurately reaches
the stationary 1+log slicing, and, since the puncture no longer represents a
second infinity, the solution is well-resolved up to the puncture. (3)  Both
the $\phi$ and $\chi$ methods are stable and accurate, but the $\chi$ method
shows (as expected) better convergence properties at the puncture. As a test of
the stability of the method at extremely high resolutions, we have also
evolved a Schwarzschild puncture with resolutions of up to $M/512$ at the
puncture, and found no signs of instability after $100M$ of evolution.

In addition, we emphasize that the only variable that diverges at the puncture
is $\phi$. When the $\chi$ method is used, {\it all} variables are finite at the
puncture. Some variables are discontinuous at the puncture. This leads to
incorrect evaluation of finite-difference derivatives at the grid points closest
 to the puncture (the number of points depends on the width of the
finite-difference stencil used), but these errors do not seem to propagate
away from the puncture, and spoil the convergence of the variables in question
only near the puncture. These errors could presumably be reduced or removed by
using appropriate one-sided derivatives next to the puncture, but we have
obtained sufficiently accurate results without need of such a sophisticated
treatment.

\subsubsection{Coordinate dependence on $\eta$}

\begin{figure}[t]
\centerline{\resizebox{9cm}{!}{\includegraphics{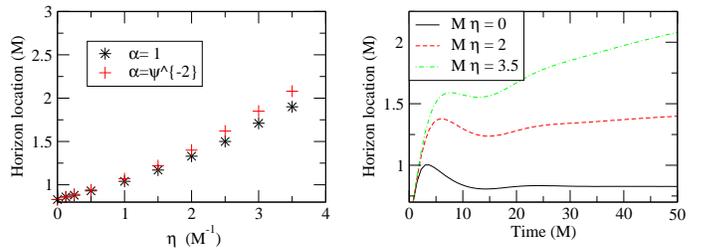}}}
\caption{Left: coordinate location of $R = 2M$ after 50$M$ of evolution,
as a function of the damping parameter, $\eta$, with initial lapse $\alpha =1$
and $\alpha = \psi^{-2}$. Right: coordinate location $R=2M$ as a function of
time, for $M \eta = 0, 2.0, 3.5$, using initial $\alpha = 1$.}
\label{fig:horizon_location}
\end{figure}

The geometry of the stationary 1+log slice is unique, but the
coordinates of that final slice are not. One quantity that alters the
final coordinates is the gamma-freezing damping parameter, $\eta$. The parameter
$\eta$ was originally introduced in \cite{Alcubierre02a} for
fixed-puncture evolutions to prevent oscillations in the shift vector
as well as long term drifts in the metric variables.
The effect of $\eta$ in our new
evolutions is demonstrated in Figure \ref{fig:horizon_location}, which shows
the coordinate location of the Schwarzschild horizon $R=2M$ after 50$M$ of
evolution, as a function of $\eta$. We see that the coordinate size of the
black hole differs by more than a factor of two between $\eta = 0$ and $\eta =
3/M$; similar effects were alluded to in \cite{Herrmann:2006ks}. 
As a result, different choices of $\eta$ correspond to different
effective numerical resolutions across the black hole. For example, with $\eta
= 0$ and a central resolution of $M/16$, there are about 26 grid points across
the interior of the black hole. With $\eta = 3/M$, there are about 59 grid
points across the black hole --- it is resolved twice as well! On the other
hand, if the finest box in the mesh-refinement structure contains $32^3$
points, then this box contains the entire black hole when $\eta = 0$, but does
not when $\eta > 1.0/M$.

In any black-hole simulation, one must decide which is more important,
the effective finest resolution, or the effective size of the finest
box. Perhaps more importantly, the effect of $\eta$ on the coordinates shows
that one must be careful when comparing runs that use different resolutions
and/or box sizes, {\it and} different values of $\eta$.

Larger values of $\eta$ also cause a larger drift in the horizon location with
time. Although the geometry becomes stationary after about $40M$ of evolution,
the numerical coordinates may not. This is clear from the lower plot in
Figure~\ref{fig:horizon_location}. We see that, if we wish to minimize the
drift in the numerical coordinates, lower values of $\eta$ are better. We will
see similar results in Section \ref{sec:Orbits} in the case of black-hole binaries.

\subsection{Numerical experiments for a single spinning puncture}
\label{SingleNumSpin}

We now look at results for evolutions of a single spinning puncture.  These
allow us to test the moving-puncture method for spinning black holes, and provide a 
non-trivial test of the wave-extraction algorithm for a black-hole
spacetime. The initial data are now based on the Bowen-York extrinsic
curvature for a single black hole with non-zero spin, which can be considered
as a Kerr black hole plus Brill-wave radiation
\cite{Bowen80,York-Piran-1982-in-Schild-lectures,Choptuik86b,Brandt94a}. In an
evolution  the additional radiation will leave the system, and only the Kerr black
hole will remain. The energy of the radiation has been estimated in the
past by studying the initial data
\cite{York-Piran-1982-in-Schild-lectures,Choptuik86b}, with a radiation
content of up to $3\%$ for a near-maximally spinning Kerr black hole.  

We considered a Bowen-York puncture with mass parameter $m_1 = 1$ and angular
momentum  parameter $S_z = 0.2m_1^2$. As discussed in Section \ref{sec:ID} the
mass of the black hole can be estimated using Eq.~(\ref{adm_punc}). For these
data, the  black-hole mass is $M = 1.0155m_1$. The Kerr parameter can then be
estimated as $a = s/M^2 = 0.194$. 

The spinning puncture was evolved for $100m_1$ using the $\phi$ and $\chi$
methods. Convergence tests consisted of runs with seven levels of refinement,
 box sizes of $40^3$, $48^3$ and $64^3$ points, and resolutions of the
 coarsest box of $h_{max} = 6m_1, 5m_1, 3.75m_1$. 

\begin{figure}[t]
\centerline{\resizebox{9cm}{!}{\includegraphics{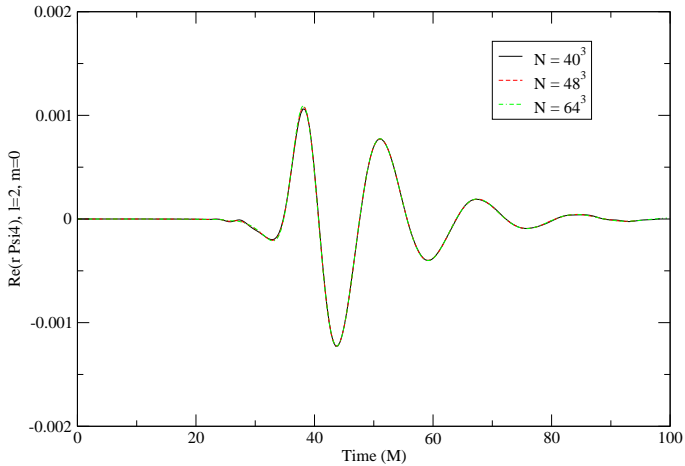}}}
\caption{Real part of the $l=2, m=0$ mode of $r\Psi_4$, extracted at $r = 30m_1$.}
\label{fig:spin_psi4}
\end{figure}

\begin{figure}[t]
\centerline{\resizebox{9cm}{!}{\includegraphics{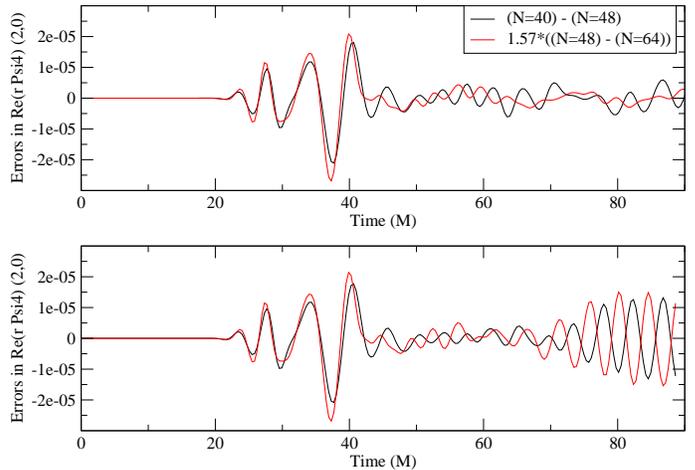}}}
\caption{Errors in the real part of the $l=2, m=0$ mode
 of $r\Psi_4$, extracted at $r = 30m_1$. The upper plot shows results from runs
 with the $\phi$ method, while the lower plots shows results with the $\chi$
 method. See text for the grid details and discussion.} 
\label{fig:spin_psi4_cvg}
\end{figure}

Figure \ref{fig:spin_psi4_cvg} shows convergence plots for the $\phi$ and
$\chi$ methods. We have plotted the differences in $Re(r\Psi_4)_{20}$ between
the three grid sizes, and scaled the medium-fine difference by a factor of
$1.57$, consistent with fourth-order convergence. Both methods show reasonable
fourth-order convergence for the first $\sim 40M$ of evolution, demonstrating
that the wave-extraction algorithm is fourth-order convergent. Convergence in
the waveform (and the evolution variables) is lost after that time. This may
be due to reflections from mesh-refinement boundaries. However, for both the
$\phi$ and $\chi$ runs the errors are extremely small, and of comparable
magnitude. 

\section{Numerical experiments for two orbiting punctures}\label{sec:Orbits}

In this section we calibrate our code for binary evolutions. These will be the
principal application of our code in future work, and we therefore perform a
more detailed study than in the case of single black holes. We
focus on runs that use the initial-data parameters of the run
``R1'' in \cite{Baker:2006yw}, for which comparison simulations were also
performed in \cite{Sperhake:2006cy}. We evolve these data with both the $\phi$
and $\chi$ variants of the moving-puncture method, and in each case compare
runs with $\eta = 1$ and $\eta = 2$, to determine which aspects of the
simulations are most strongly affected by different gauge choices. 

The main goal of this paper is to demonstrate the accuracy and efficiency of
our code, and of course to verify that it gives correct results. We begin by
determining the grid setups necessary to achieve fourth-order-accurate
results, and present our results with error bars calculated from the
difference between the highest-resolution runs and Richardson extrapolated
values. By ``grid setup'' we do not simply mean ``resolution''; the sizes of
the mesh-refinement boxes are also important, both for the accuracy of the
simulation, and the extracted physical quantities. Having done that, we
compare our extracted waveforms with those produced by the independent LEAN
code \cite{Sperhake:2006cy}. This is an extremely strong test: it validates both
codes, and also demonstrates the high accuracy of their results. Finally, we
study in detail the accuracy of various extracted physical quantities
(radiated energy, angular momentum, and angular frequency during inspiral),
and their dependence on the radiation extraction radius and the gauge
parameter $\eta$.

\subsection{Setup}

The initial-data parameters for the runs in this section are: the
punctures have mass parameters $m_1 = m_2 = 0.483$, and are placed on
the $y$-axis at $y = \pm 3.257$ with momenta $p_x = \mp 0.133$. The
individual black-hole masses, as determined by Eq.~(\ref{adm_punc}),
are $M_1 = M_2 = 0.505$, and the total ADM mass of the spacetime is
$E_{ADM} = 0.996$. These parameters correspond to the run ``R1''
in~\cite{Baker:2006yw}. They result in $\sim 1.8$ orbits before merger
at roughly $160M$. We define three times indicating the merger time:
$t_{AH}$, the time when an apparent horizon first forms, $t_\alpha$,
the time at which the lapse at the center drops below the value $0.3$
(following e.g.,~\cite{Baker:2006yw,Reimann:2003zd}), and
$t_{max}$, the time at which $\vert\Psi_4\vert$ reaches a maximum
(which depends on the extraction radius $r_{ext}$). While $t_{AH}$ is
of immediate relevance regarding the simulation, it is also more
costly to evaluate accurately, while accurate evaluation is trivial
for $t_\alpha$ and $t_{max}$. We therefore find it very useful to
check convergence in phase by evaluating $t_\alpha$ and $t_{max}$,
and note that $t_\alpha$ and $(t_{max} - r_{ext})$ give an estimate
for $t_{AH}$ which is accurate to a few $M$.

Note that in this section, all distances and
times are either scaled with respect to the total black-hole mass, $M$ (consistent
with our discussion of single black holes in Sec.~\ref{sec:Single}), in
which case the appropriate unit is given (e.g., $r \Psi_4 (M^{-1})$), or, when
no rescaling has been done, the numerical coordinates are used (e.g.,
``extraction at $r=30$'').

We label the grid setups for orbit runs with the notation
$X[n_1\times N_1:n_2\times N_2:{buf}][h_{min}^{-1}:h_{max}]$,
where $X$ denotes the choice of conformal factor $\phi$ or $\chi$,
and the grid is composed of $n_1$ levels of $N_1^3$ grid points and
$n_2$ levels of $N_2^3$ grid points (reducing the number of grid points
appropriately when discrete symmetries are applied), and $buf$ mesh
refinement buffer points are used (not counting ghost zones for
parallelization). The quantities $h_{min}$ and $h_{max}$ denote the grid
spacing on the finest and coarsest levels.  The qualifier $X$ occasionally
carries subscripts specifying further parameters. 
Examples would be $\phi[5\times32:5\times64:6][38.4:8]$
or $\chi_{\eta=0.05}[5\times32:5\times64:6][38.4:8]$.
The ratio of grid spacings between neighboring levels is always two. 

We have performed a large number of runs, both complete convergence series and
lower resolution ``exploration runs'' with different grid layouts, gauges or 
numerical methods --- for the presentation here we have to make a selection
and present results from three series of runs, which we have found typical: 
\begin{enumerate}
\item BAM$\phi$1: $\phi_{\eta=1}[4\times{}i:6\times2i:6]$. 
       
\item BAM$\chi$1: $\chi_{\eta=1}[5\times{}i:4\times2i:6]$. 

\item BAM$\chi$2: $\chi_{\eta=2}[5\times{}i:4\times2i:6]$, i.e., as above, but with $\eta=2$. 
\end{enumerate} 
The BAM$\phi$1 series is representative of our early experiments. Apart from
using the $\phi$ evolution variable, they also used the $ttt$ gauge advection choice.
We later found the BAM$\chi$ runs to be more accurate. In addition,
the merger times converged from below, and convergence behavior was monotonic
even at low resolutions. 
For the runs presented here, we see no strong difference between the $ttt$ and
$000$ gauge advection choices as described in Section \ref{Gauge}. However. the
manifestly strong hyperbolicity of the $000$ gauge \cite{Gundlach:2006tw}
makes that choice more attractive.  For the $ttt$ choice, the slow-speed modes
described in \cite{vanMeter:2006vi} can be clearly seen in animations of the
grid variables. Both choices yield stable evolutions, but we regard the $000$
choice as superior and have used it in the BAM$\chi$ runs presented here and
in subsequent work. 

The runs presented here have 9 and 10 refinement levels (labeled from 0 for 
the coarsest to 8 or 9), and use twice the number
of grid points on the outer levels than on the finest levels, as detailed in
Table \ref{tab:orbit_grids_BAM}.  We find that this setup yields 
higher accuracy for wave extraction without too drastic an increase in
computational cost. 
Typical performance numbers of our code are displayed in Table
\ref{tab:timing}. All runs are carried out with the symmetry $(x,y,z)
\rightarrow (-x,-y,z)$ and $(x,y,z) \rightarrow (x,y,-z)$, reducing the
computational cost by a factor of four compared with runs that do not exploit
any discrete symmetries. The courant factor $C=\Delta t/h_l$ is kept constant,
and is set to $C=1/2$ for the inner grids, while for the outer grids at levels
0--2 the time step  is kept constant at the value of level 3. All runs
presented here use the RK4 time integration scheme. Using the ICN scheme
(without artificial dissipation) did not change results significantly. We find
that for a constant courant factor we occasionally encounter numerical
instabilities in the outer regions of the simulation domain, but these were
cured by freezing the size of the time step in the outermost 3 (BAM$\chi$
runs) or 4 (BAM$\phi$1 run) levels. 

All the BAM runs presented here use six AMR buffer points (see
Sec.\ \ref{sec:Numerics}), which is less than required to isolate the fine
level ``half'' timestep from time interpolation errors at the mesh-refinement
boundary, and in particular also less than required for the fully fourth-order
Christmas-tree scheme suggested in \cite{Lehner:2005vc}. We have experimented with
using higher numbers of buffer points up to the number required for the
Christmas-tree scheme, but have not found significant improvements in the
results, which is consistent with the fact that we find fourth-order
convergence and no significant improvement of the results when decreasing the
timestep. We conclude that at the resolutions presented here, six buffer points
are enough to suppress errors from interpolation in time at mesh-refinement
boundaries below the relevant threshold as far as the dynamics and low
frequency waves are concerned. To suppress high-frequency reflections at the
mesh-refinement boundaries, which can we have seen in quantities like
$\Psi_4$ or the constraints, we use fourth-order Kreiss-Oliger dissipation as
described in Section~\ref{sec:Numerics}, where the factor $\sigma$ is chosen
as $\sigma = 0.1$ in the inner levels and  $\sigma = 0.5$ in the outer levels
(where the waves are extracted). 

\begin{table}
\begin{ruledtabular}
\begin{tabular}{l|r|r|r}
Run & $h_{min}$  & $h_{max}$ & $r_{max}$  \\[10pt]
\hline
$\phi_{\eta=1}[4\times 32:6\times 64:6]  $ & $1/25.6$ &  20   & 648.0\\
$\phi_{\eta=1}[4\times 40:6\times 80:6]  $ & $1/32.0$ &  16   & 672.0\\
$\phi_{\eta=1}[4\times 48:6\times 96:6]  $ & $1/38.4$ &  40/3 & 666.7\\
$\phi_{\eta=1}[4\times 64:6\times 128:6] $ & $1/51.2$ &  10   & 680.0\\
$\phi_{\eta=1}[4\times 72:6\times 144:6] $ & $1/57.6$ &  80/9 & 644.4\\
\hline
\hline
$\chi_{\eta=1,2}[5\times 32:4\times 64:6]  $ & $1/25.6$ &  10   & 325.0 \\
$\chi_{\eta=1,2}[5\times 40:4\times 80:6]  $ & $1/32.0$ &  8    & 324.0 \\
$\chi_{\eta=1,2}[5\times 48:4\times 96:6]  $ & $1/38.4$ &  20/3 & 323.3 \\
$\chi_{\eta=1,2}[5\times 56:4\times 112:6] $ & $1/44.8$ &  40/7 & 322.8 \\
$\chi_{\eta=1,2}[5\times 64:4\times 128:6] $ & $1/51.2$ &  5    & 322.5 \\
$\chi_{\eta=1,2}[5\times 72:4\times 144:6] $ & $1/57.6$ &  40/9 & 322.2 \\
\end{tabular}
\end{ruledtabular}
\caption{\label{tab:orbit_grids_BAM}Grid setups used for binary
  evolutions. See text for definition of the notation in the ``Run''
  column. $h_{min}$ and $h_{max}$ (rounded to 3 digits) denote the finest and
  coarsest grid spacings, and $r_{max}$ is the location of the outer boundary
  (rounded to 4 digits).}
\end{table}
\begin{table}
\begin{ruledtabular}
\begin{tabular}{l|r|r|r|r}
grid configuration &  procs.  & mem. (GByte)  & time & M/hour\\[10pt]
\hline
$\chi[5\times 56:4\times 112:6]$   & 10 &  8.9 & 192 & 18.2 \\
$\chi[5\times 64:4\times 128:6]$   & 12 & 11.8 & 306 & 13.7 \\
$\chi[5\times 72:4\times 144:6]$   & 14 & 17.5 & 505 &  9.7 
\end{tabular}
\caption{\label{tab:timing}Typical performance results for runs lasting $350
  M$: number of processors, maximal memory requirement in GByte (to be
precise, we quote the resident size of the program, i.e.~the physical memory a task has used), total runtime
  in CPU hours and average speed in $M$/hour for the Altix 4700  of LRZ
  Munich \cite{hlrb2} (using Intel Itanium2 Madison 9M CPUs running at 1.6 GHz).  }
\end{ruledtabular}
\end{table}

\subsection{Results}

\begin{figure}[ht]
\centerline{\resizebox{9cm}{!}{
\includegraphics{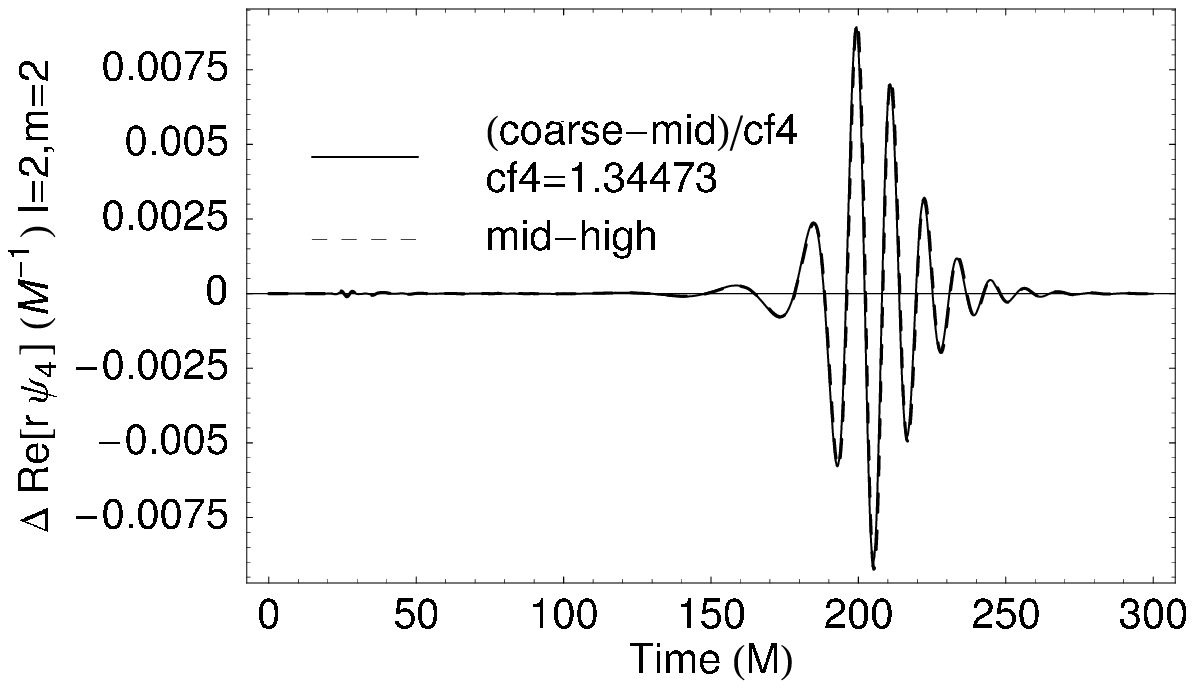}
}}
\centerline{\resizebox{8cm}{!}{
\includegraphics{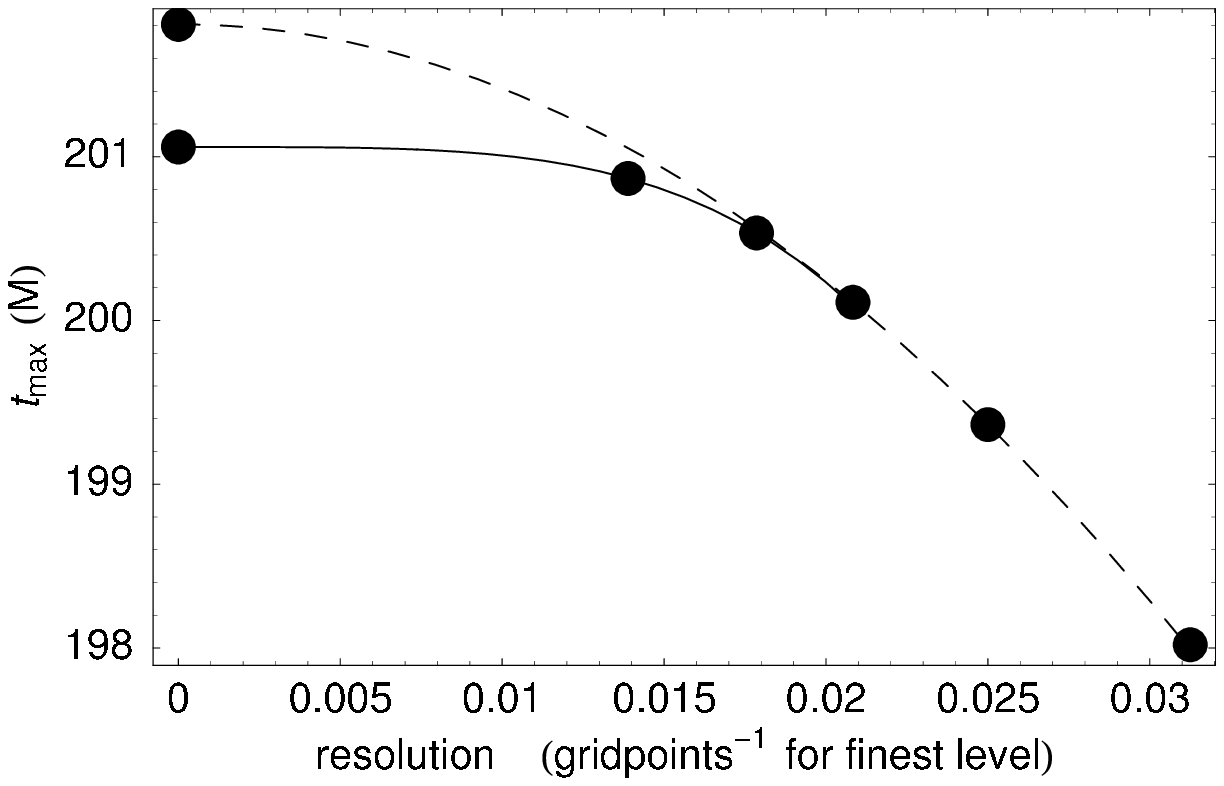}
}}
\caption{
Convergence of the BAM $\chi_{\eta=2}[5\times i:4\times 2 i:6][i=32,40,48,56,72]$
series. 
Top image: scaled differences between results for different resolutions
demonstrate fourth-order convergence.
Bottom image: the time $t_{max}$ (in units of $M$) of the maximal amplitude in $\Psi4$ is
plotted versus resolution, and Richardson extrapolated values are shown. The
three most accurate runs ($i=48$, $56$ and $72$) show fourth-order
convergence, while the runs with ($i=32$, $40$, $48$ and $56$) show
second-order convergence.} 
\label{fig:chi2_wave_convergence}
\end{figure}

We have obtained fourth-order convergence for $r\Psi_4$ and the
puncture tracks for sufficiently high resolution in the BAM$\phi$1,
BAM$\chi$1 and BAM$\chi$2 series, requiring at least $48$ grid points
on the fine levels. For the BAM$\phi$1 series, low resolutions, e.g.,
with $\phi_{\eta=1}[4\times 32:6\times 64:6]$, $i=32,40,48$, show no
systematic convergence behavior, while for the $\chi$ series, low
resolutions show a convergence behavior as illustrated in the lower
panel of Fig.\ \ref{fig:chi2_wave_convergence}: convergence is at
least monotonic, but only the runs with high resolution are consistent
with fourth-order convergence.

\begin{figure}[ht]
\centerline{\resizebox{9cm}{!}{
\includegraphics{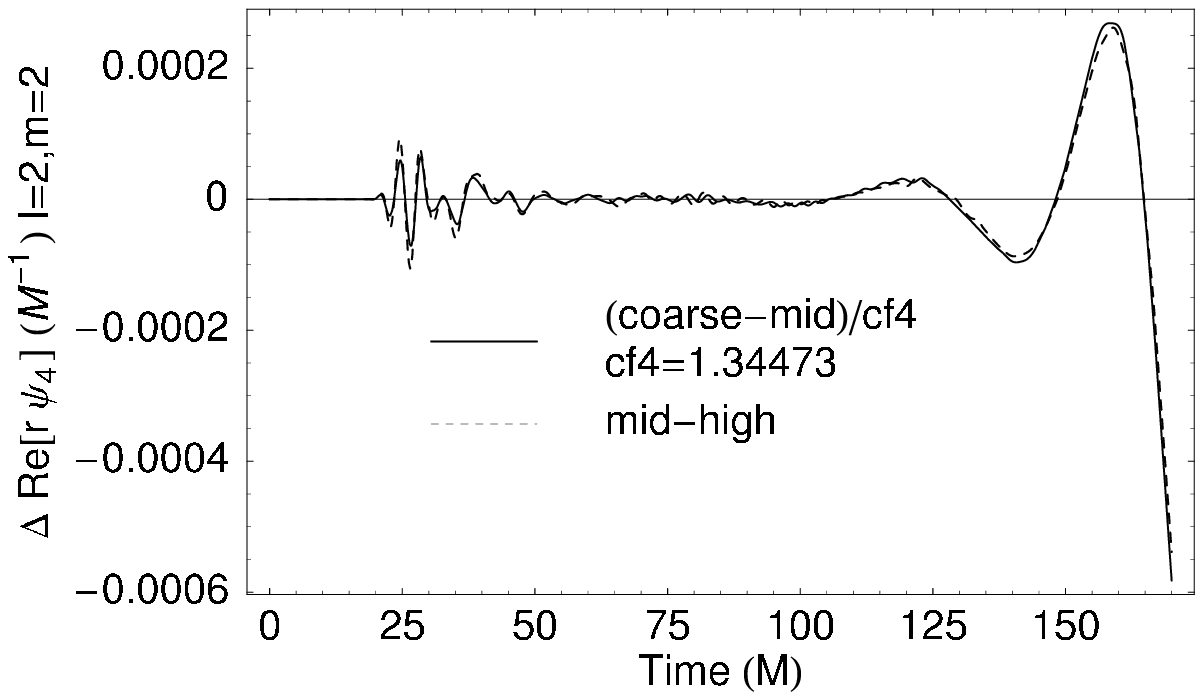}
}}
\centerline{\resizebox{9cm}{!}{
\includegraphics{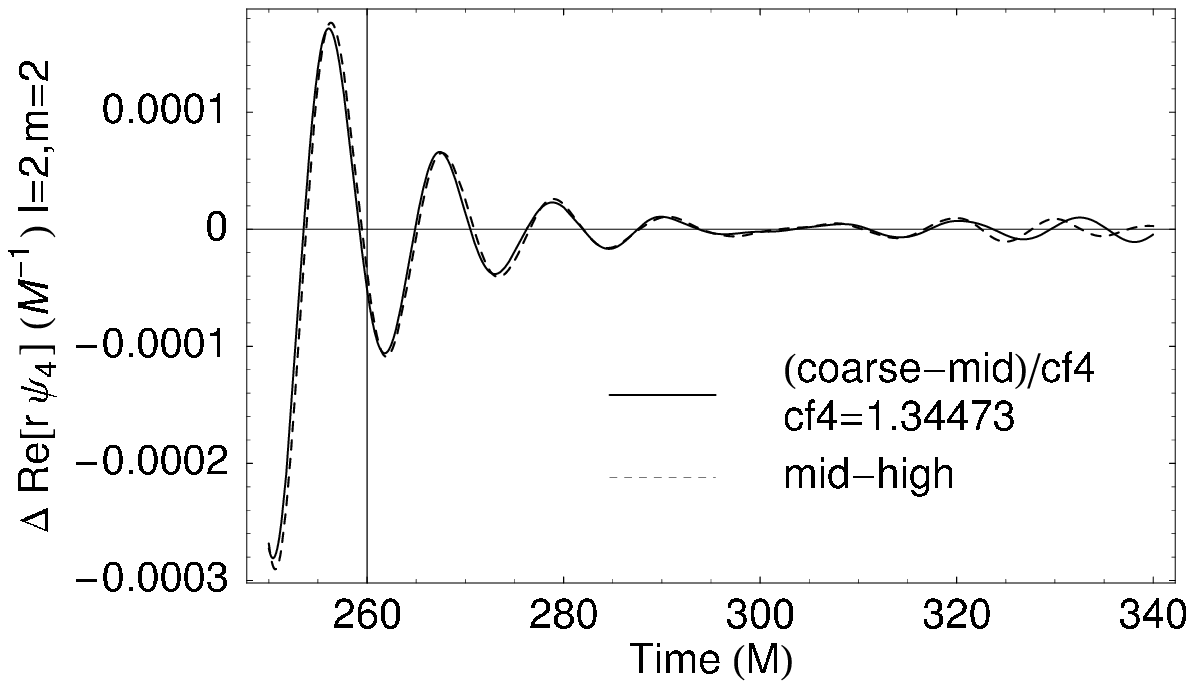}
}}
\caption{
Convergence of the BAM $\chi_{\eta=2}[5\times i:4\times 2 i:6][i=48,56,72]$ series.
Upper image: the early part of the waveform does not show fourth-order convergence until approximately
$t=125 M$.
Lower image: clean fourth-order convergence is lost at approximately $t=310 M$.
}
\label{fig:chi2_wave_convergence_intermediate_late}
\end{figure}

We now focus on the BAM$\chi$2 series.
The runs $\chi_{\eta=2}[5\times{}i:4\times2i:6]$ show approximately
second-order convergence for the set $i = 32,40,48$, but achieve clean
fourth-order convergence with $i = 48,56,72$. The convergence results are
shown in Figures~\ref{fig:chi2_wave_convergence} and
\ref{fig:chi2_wave_convergence_intermediate_late}. Note that we see convergence
in the waveforms {\it without} the time shift performed in
\cite{Baker:2006yw}. In particular, the bottom plot in
Figure~\ref{fig:chi2_wave_convergence} shows the convergence
of the time of the maximum amplitude in $r \Psi_4$ (extracted at $40M$), as a
function of the resolution on the finest box. We see that $t_{max} = 204.65
\pm 1.3 M$ for the second-order-accurate results, and $t_{max} = 203.9 \pm
0.2M$ for the fourth-order accurate results. These results are extremely
accurate; compare, for example, with the results in
\cite{Campanelli:2006gf,Baker:2006yw}.  

\begin{figure}[ht]
\centerline{\resizebox{9cm}{!}{
\includegraphics{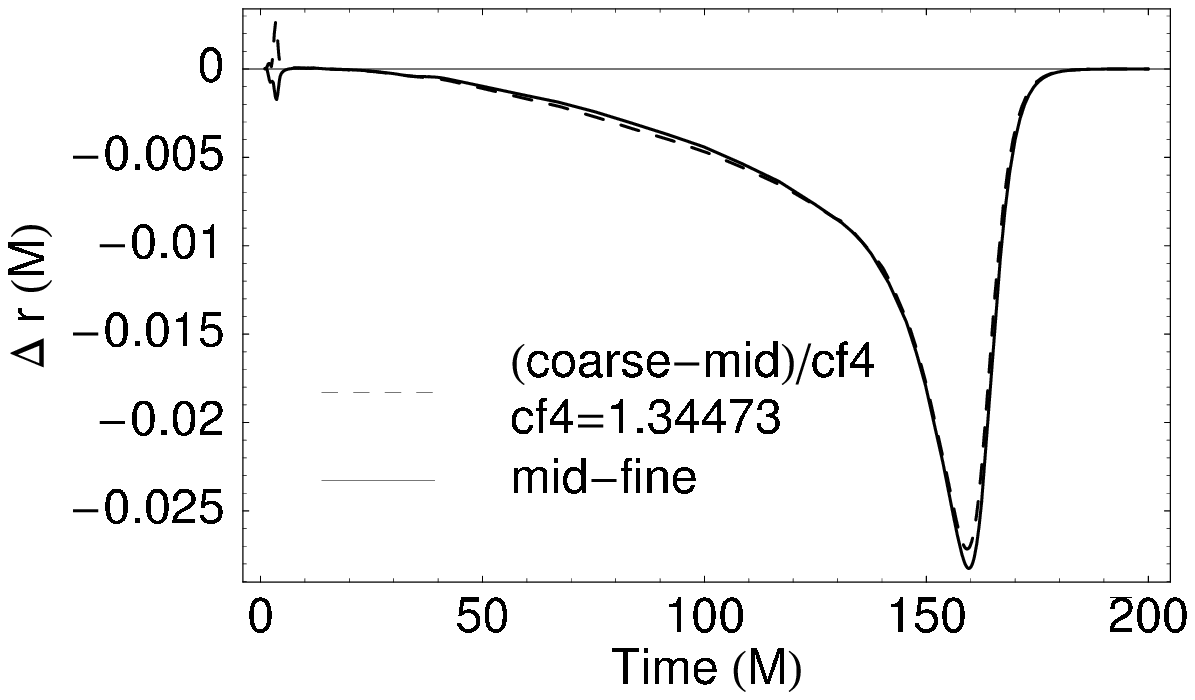}
\includegraphics{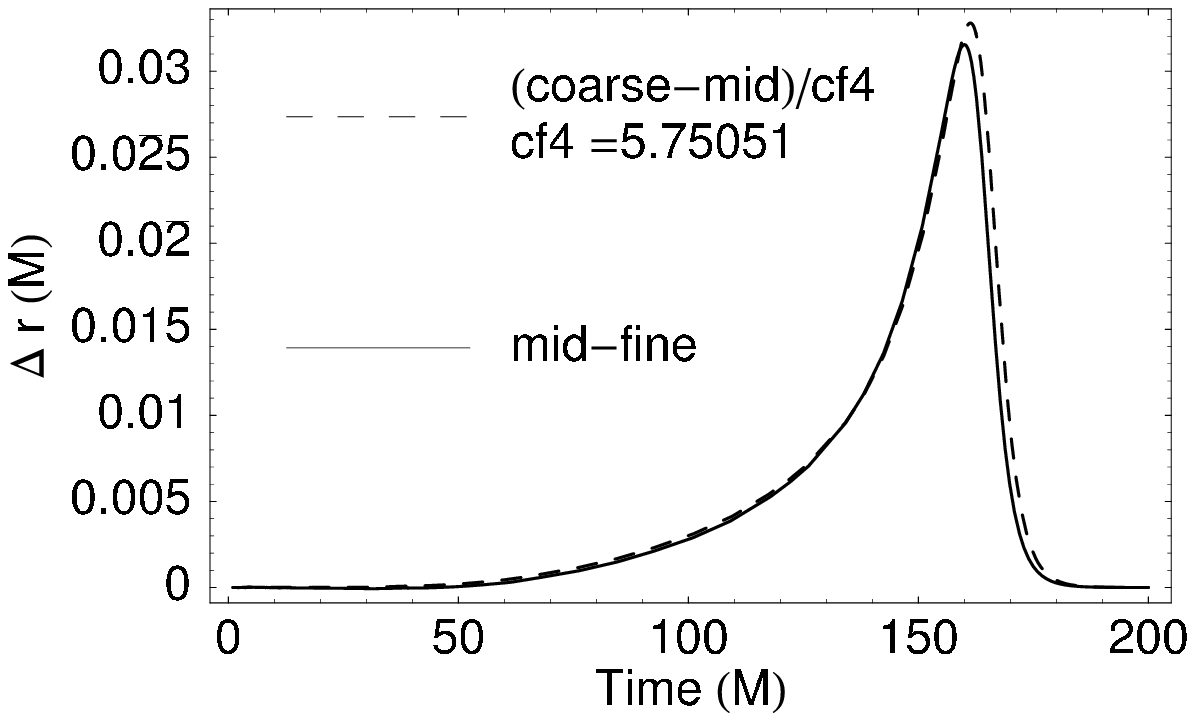}
}}
\centerline{\resizebox{9cm}{!}{
\includegraphics{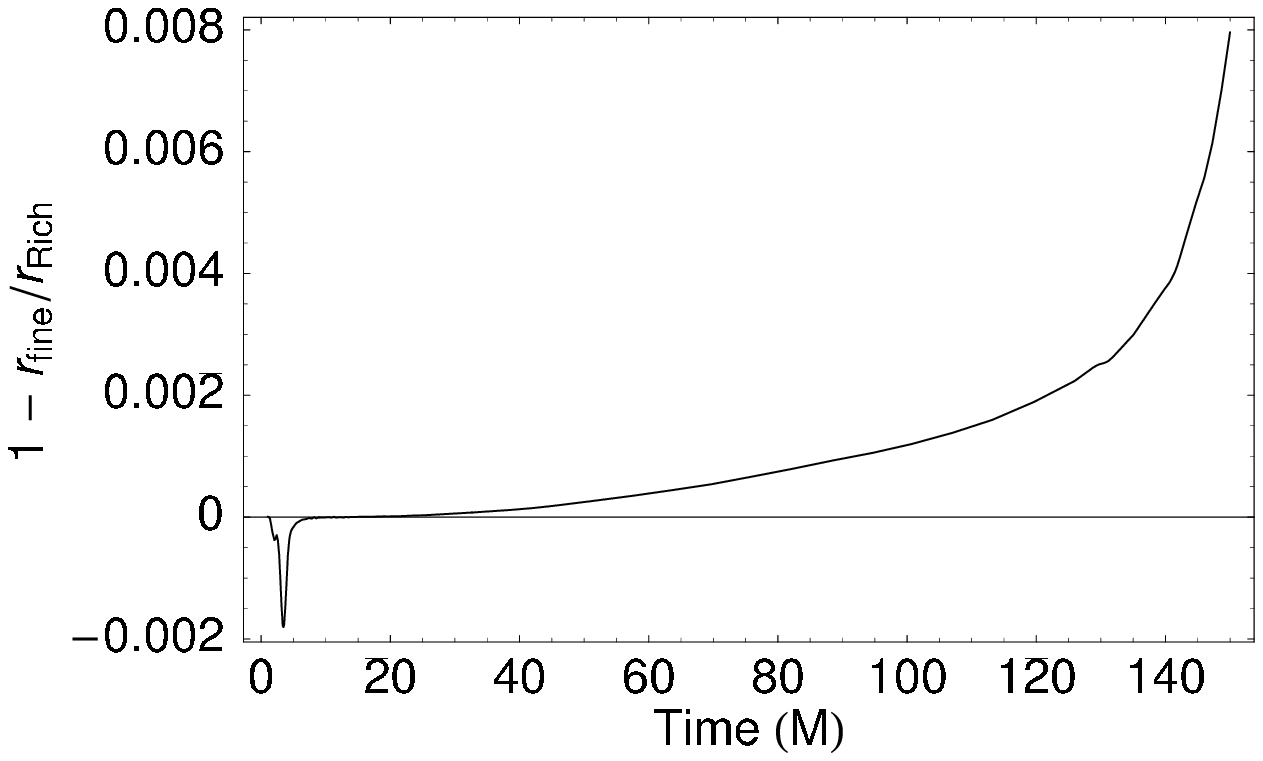}
\includegraphics{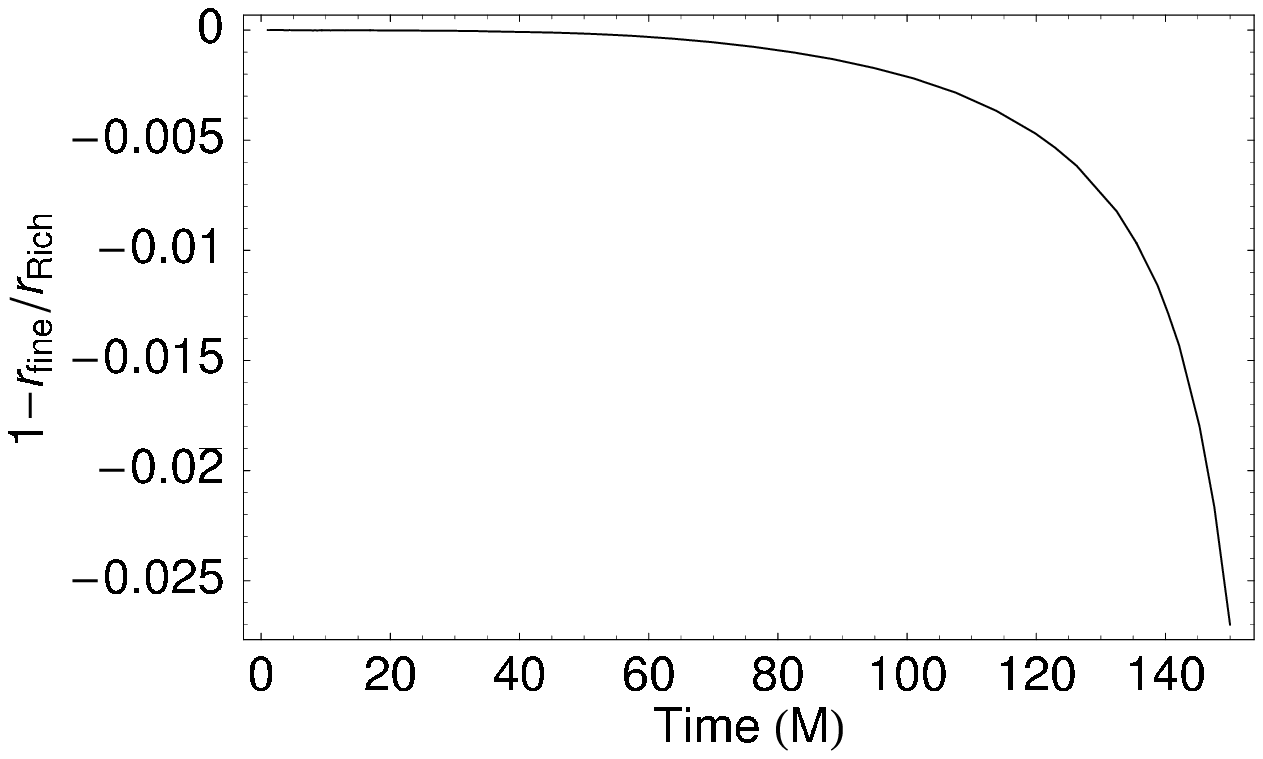}
}}
\caption{Fourth order convergence for $r(t)$ demonstrated for $\chi_{\eta=2}$
  and $\phi_{\eta=1}$ series (left and right upper images). Lower images:
  relative errors obtained from difference of Richardson extrapolated value to
  highest resolution result (left:  $\chi_{\eta=2}$, right: $\phi_{\eta=1}$).
}
\label{fig:r_convergence}
\end{figure}

\begin{figure}[ht]
\centerline{\resizebox{9cm}{!}{
\includegraphics{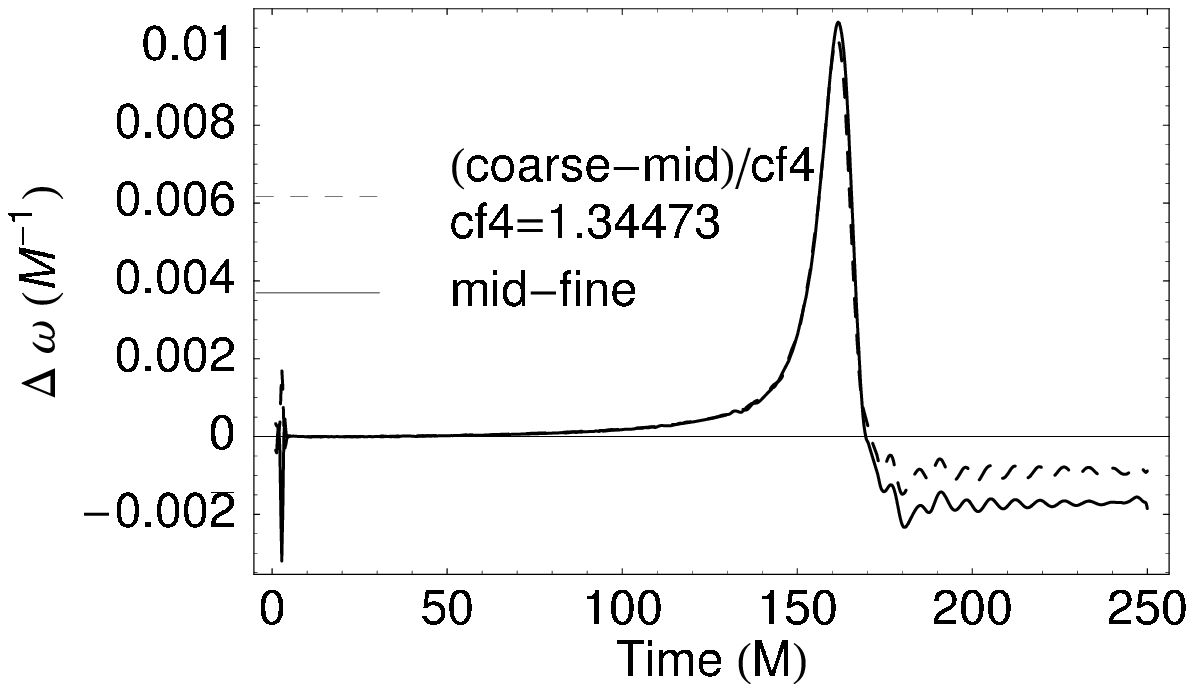}
\includegraphics{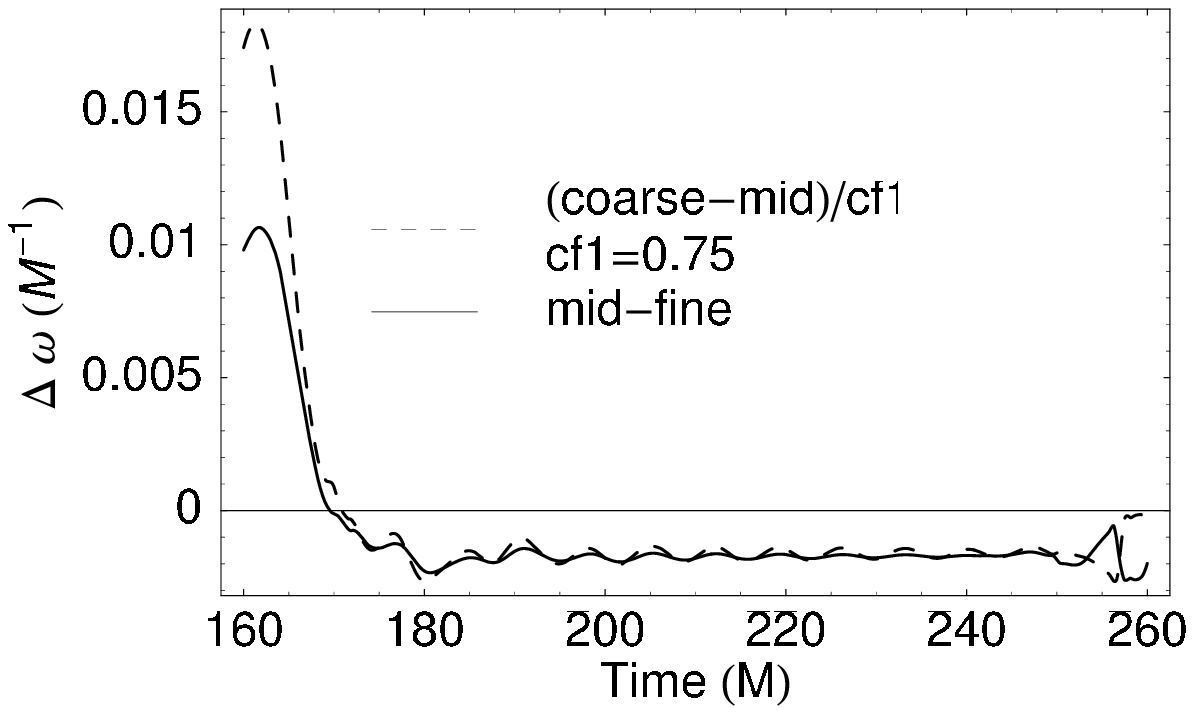}}}
\caption{Fourth order convergence for $\omega(t)$ demonstrated for
  $\chi_{\eta=2}$ series, at late times only first order convergence is seen.} 
\label{fig:w_convergence_chi2}
\end{figure}

\begin{figure}[ht]
\centerline{\resizebox{9cm}{!}{
\includegraphics{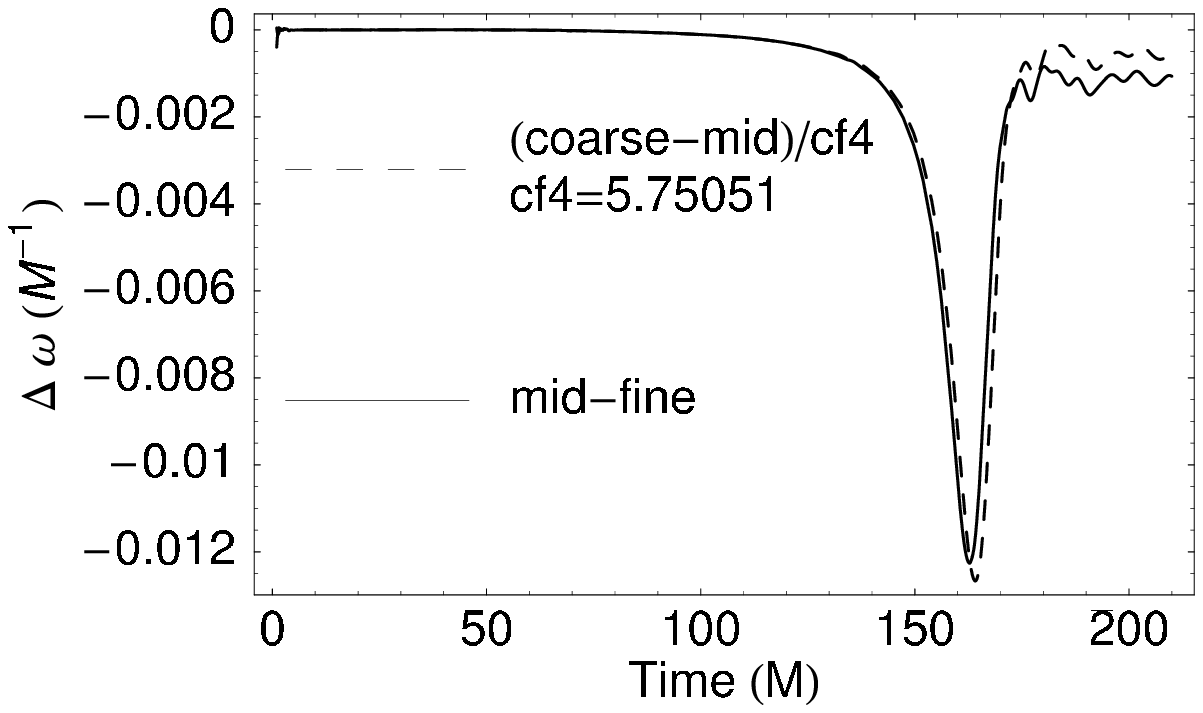}
\includegraphics{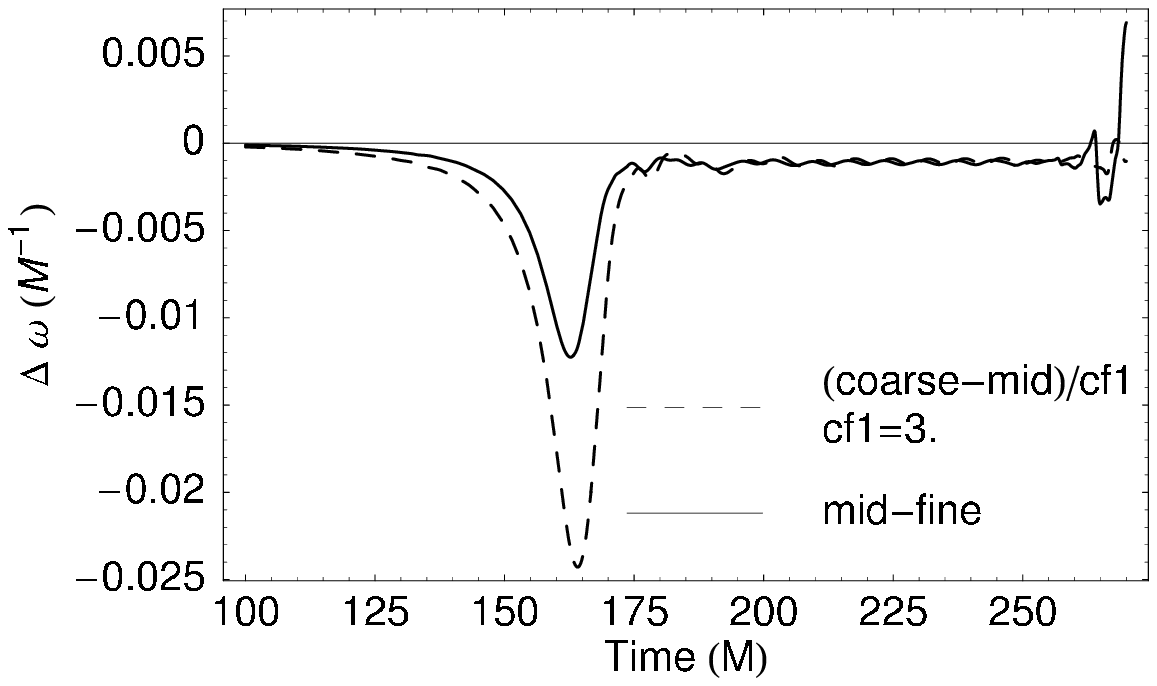}}}
\caption{Fourth order convergence for $\omega(t)$ demonstrated for
  $\phi_{\eta=1}$ series, at late times only first order convergence is seen.} 
\label{fig:w_convergence_phi1}
\end{figure}

A few comments about these results are in order. One might have stopped at
the second-order convergent $i=32,40,48$ results indicated in the lower panel 
of  Fig.\ \ref{fig:chi2_wave_convergence} for the BAM$\chi$2-runs,
since the code {\em does} have second-order ingredients (the initial data
calculation and interpolation in time at mesh-refinement boundaries). However,
this theory is vetoed by finding that neither the accuracy of the initial data,
nor the number of mesh-refinement buffer points, nor the size of the time step,
have a significant influence on the result. In such a situation it is
necessary to increase the resolution in convergence tests, and indeed for
higher resolutions fourth-order convergence was found. Note also that fourth-order
convergence does not extend to the early (before approximately $t=125 M$) and
very late (after approximately $t=310 M$) parts of the waveform, where the
errors are very small (Fig.\ \ref{fig:chi2_wave_convergence}), and to the late
part of the puncture tracks after the merger as shown in
Figs.\ \ref{fig:r_convergence}, \ref{fig:w_convergence_chi2} and
\ref{fig:w_convergence_phi1}. The late-time loss of convergence may be due to
the location of the outer boundary, which is at $\sim 320M$, and we expect
incoming errors to reach the extraction radius (at $40M$) after about
$280M$. It is encouraging that these errors are so small that they are
detectable only in a convergence test, and 
that they do not have any effect on the stability of the runs.


\begin{figure}[ht]
\centerline{\resizebox{9cm}{!}{\includegraphics{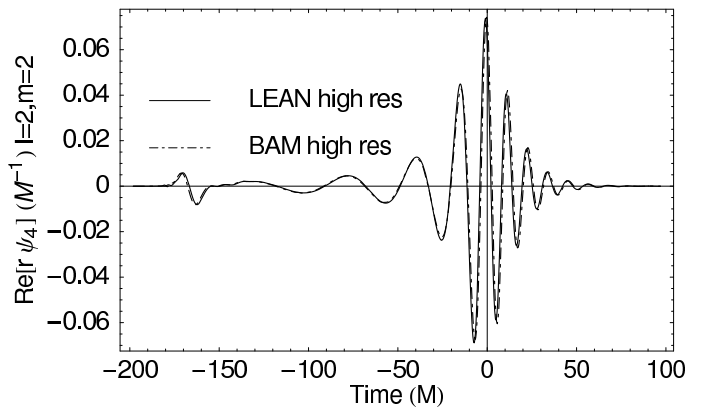}}}
\caption{
Overlay of $Re(r\Psi_4)$, extracted at $r=30$ for highest resolution
LEAN $\phi_{\eta=1}$ (as published in \cite{Sperhake:2006cy}) 
and BAM $\chi_{\eta=2}[5\times 72:4\times 144:6][44.8,5.714]$ runs.
The results have been shifted in time so $\max(|\Psi4|)$ is aligned with
$t=0$.
}
\label{fig:BAM_LEAN_waves}
\end{figure}

\begin{figure}[ht]
\centerline{\resizebox{9cm}{!}{\includegraphics{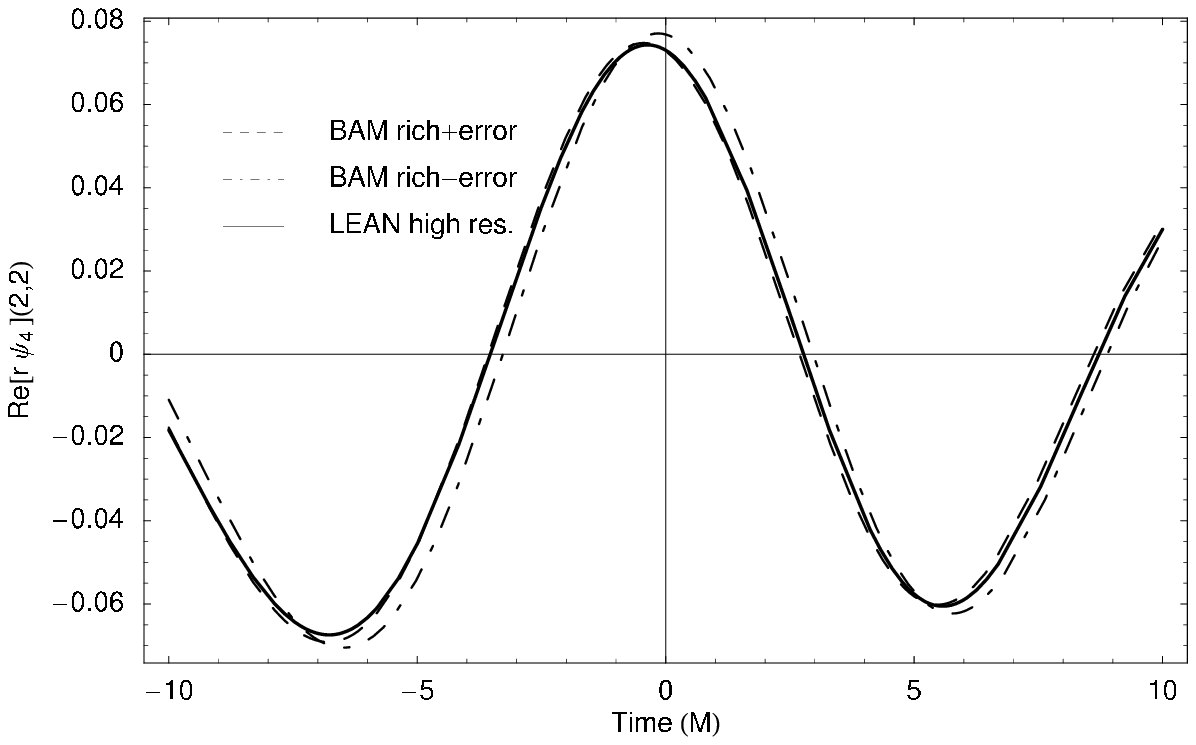}}}
\caption{
The Richardson-extrapolated ``main'' part of the waveform is shown for the BAM $\chi_{\eta=2}$ 
runs, error bars are obtained from the difference between Richardson-extrapolated result and 
the highest resolution run. 
The BAM result is overlaid with the highest resolution LEAN result.
The results have been shifted in time so $\max(|\Psi4|)$ is aligned with
$t=0$.
}
\label{fig:BAM_LEAN_mainwaves}
\end{figure}

Having established fourth-order convergence of our runs in the given
regime, and having determined a grid and parameter setup that produces
accurate results, we now perform an independent validation of our
results by comparing the $\Psi_4$ waveforms with those obtained with
the LEAN code, as published in~\cite{Sperhake:2006cy}. In the notation
introduced above, the grid specifications of the high-resolution LEAN
code run are $\phi_{\eta=1}[2\times72:6\times130:3][32.92:3.47]$. This
comparison is shown in Figs.~\ref{fig:BAM_LEAN_waves} and
\ref{fig:BAM_LEAN_mainwaves}. Figure~\ref{fig:BAM_LEAN_waves} shows
highest-resolution BAM and LEAN $r\Psi_4, l=2, m=2$ modes, extracted at $r =
30$, for full runs. Figure~\ref{fig:BAM_LEAN_waves} focuses on the main
part of the waveform, and error bars are shown for the BAM results. We see
that the results from the two codes show excellent agreement. 


\begin{figure}[ht]
\centerline{\resizebox{9cm}{!}{
\includegraphics{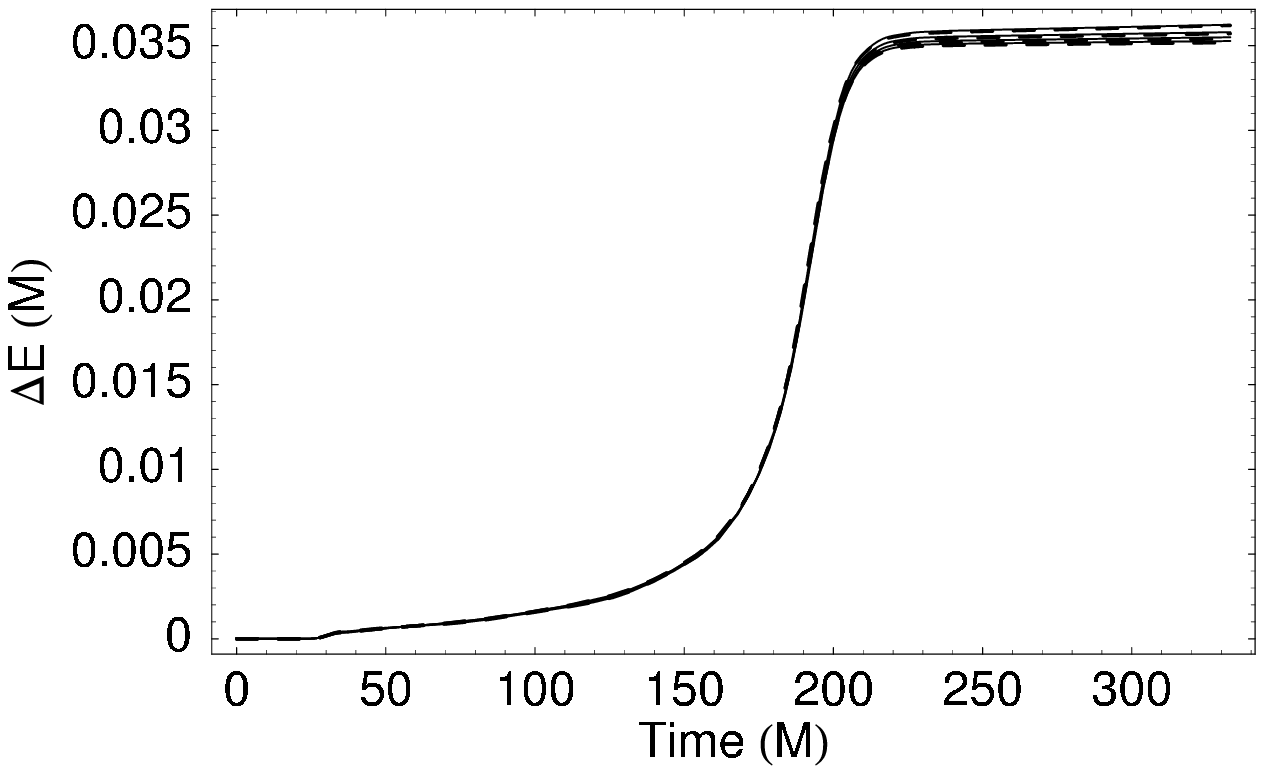}
\includegraphics{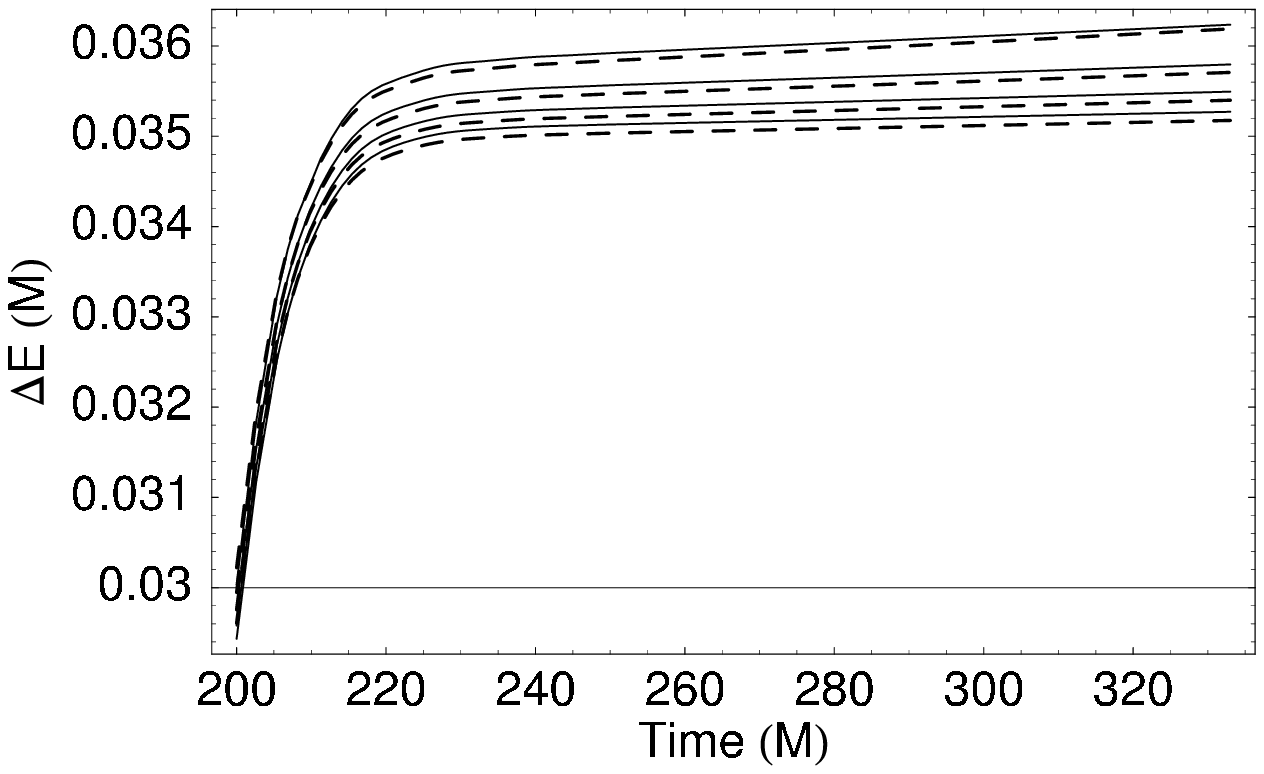}
}}
\caption{
Radiated energies plotted versus time for the
runs $\chi_{\eta=2}[5\times i:4\times 2 i:6]$ ($i=56,72$)  and
extraction radii $r=25,30,35,40$. The dashed lines are from the $i=56$ run,
the full lines from the $i=72$ run. Larger extraction radii yield smaller
values for the radiated energy. The left image shows the result for the
complete run, the right images zooms in on the late stage of the run. Results
have been shifted in time by Eq.~(\ref{eq:Tortoise}) to minimize the 
phase difference. Clearly, the error from the variation of extraction radius 
is not smaller than numerical error.
}
\label{fig:energy_vs_radius}
\end{figure}

The main focus of this paper is to present and validate our code, and show
that we are able to produce highly accurate results with moderate
computational resources. In this spirit we present numerical error bars for
our results, which are easily determined from the difference between
highest resolution runs and Richardson extrapolated values. However, waveforms
and radiated energies also come with errors due to the finite extraction
radius of the waves, and due to physically inappropriate boundary
data.
Fig.~\ref{fig:energy_vs_radius} shows radiated energies
versus time for $\chi_{\eta=2}$ BAM runs
$\chi_{\eta=2}[5\times i:4\times 2 i:6]$, ($i=56,72$) 
and
extraction radii $r=25,30,35,40$. The dashed lines are from the
$i = 56$ run and the full lines from the $i=72$ run. The results have been
shifted in time by the differences of extraction radii to minimize the phase
difference. Clearly, the error from the variation of extraction radius is
larger than numerical error at radii less than $30M$. Assuming that the error
falls off with some power of $r$, a curve fit of our results suggests that the
error falls off as $r^2$, and that in the $r \rightarrow \infty$ limit the
extracted energy is $3.52$\%. The value extracted at $r=40$ therefore has an
error of only about 2\%, in contrast to the numerical error, which is less
than 0.3\%. Further progress in accuracy obviously makes it very desirable
to better model the fall-off properties of the radiation. 
Although  the only completely aesthetically pleasing solution would be
to compactify at null infinity (see 
\cite{Winicour98,Husa:2002kk,Husa02a,Andersson:2002gn,Babiuc:2005pg,Husa:2005ns,Reisswig:2006nt} for some
recent work and overviews), one should expect that simple estimates based on perturbations of
Kerr can be used to obtain significant improvements. A more detailed analysis goes beyond the
scope of this paper, which focuses on the numerics, and will be published elsewhere.

The idea of extracting wave signals at finite distance from the source
is that the timelike cylinder traced out by a sphere at sufficiently
large distance from the source can be viewed as an approximation of
null infinity, and increasing the distance will increase the quality
of the approximation. Thus an approximation to the signal expected at
a detector located at an astronomical distance from the source can be
calculated. When the distance is of cosmological scale, cosmological
redshift or further effects would have to be added ``by hand''. This
idea raises several serious issues: the error introduced by the cutoff
at finite radius needs to be estimated, ``extrapolation procedures''
to larger radius may yield significant improvements if a fall-off law
for the finite distance results can be assumed, and the gauge
dependence of the results at finite distance needs to be addressed in
order to optimize such procedures.  It is therefore very valuable to
compute results characterizing the asymptotics of the gravitational
field by different methods, which may show different effects from
gauge, or different fall-off laws, and compare the results of such
different prescriptions of asymptotic quantities.

\begin{figure}[t]
\centerline{\resizebox{7cm}{!}{\includegraphics{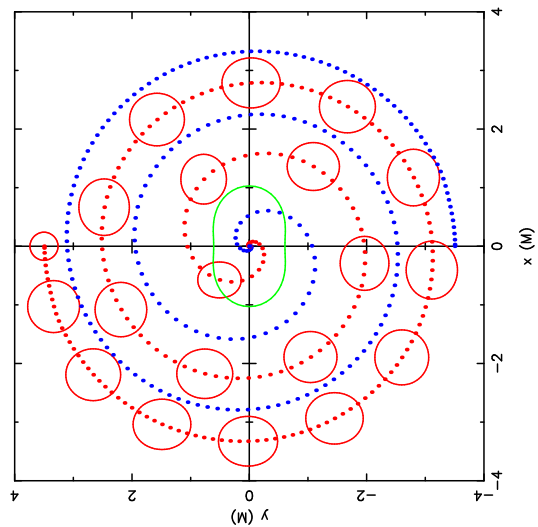}}}
\caption{Shape of the apparent horizon in the x-y plane plotted each 10M 
for a simulation with initial separation of $r=3.5$M. The dotted line
represents the trajectory of the puncture. The behavior of the second
horizon can be obtained by the symmetry of the problem. The common
apparent horizon (peanut-shaped in the figure) appears when the black
holes have merged.}
\label{fig:orbits_with_horizon}
\end{figure}

\begin{figure}[t]
\centerline{\resizebox{7cm}{!}{\includegraphics[keepaspectratio]{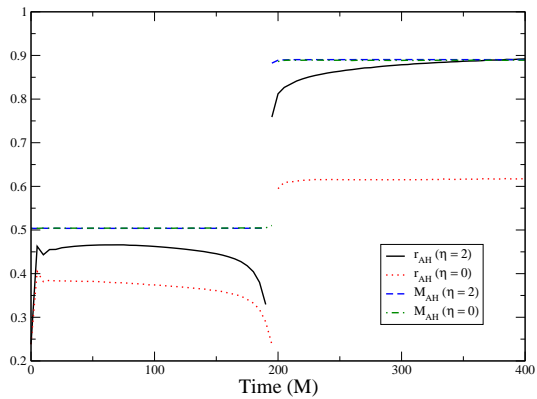}}}
\caption{Coordinate radius $r_{AH}$ of the apparent horizons as function of time for 
$\eta = 0$ and $\eta = 2$. The choice of $\eta$ has a strong influence in
this gauge dependent quantity. The apparent-horizon mass $M_{AH}$ does not
show any such gauge dependence.} 
\label{fig:horizon_r_and_m}
\end{figure}

Along these lines, a good check on the consistency of the wave extraction
algorithm is to compare radiated energies with the energy balance that can be
determined from evaluating the mass integral Eq.~(\ref{madm_int}) 
at the beginning and end of the integration time. At finite extraction radius
one can determine an estimate for the difference in Bondi mass; see Section
\ref{sec:ADM}. We have done this for all of our runs, and find excellent
agreement for $\eta=1$, and less accurate results (approximately $4\%$ of
radiated energy, i.e., roughly 10\% error) for 
$\eta=2$. The poor results for $\eta=2$ may be due
a drift in the coordinates. Such a drift was seen in the coordinate radius of
the horizon in the Schwarzschild case in Section \ref{SingleNum}. For the case
of orbiting 
punctures, we locate apparent horizons using a horizon finder based on the
{\tt AHFinder} code in the Cactus infrastructure \cite{Alcubierre98b}. Figures
\ref{fig:orbits_with_horizon} and \ref{fig:horizon_r_and_m} show the motion of
the apparent horizons for an orbital evolution (with punctures evolved from
initial positions $y=\pm 3.5$), and also the coordinate radius of the single,
and eventually common, apparent horizons as a function of time. We once again
see an $\eta$-dependent coordinate drift, which we expect also affects the
quality of the Bondi mass. However, note that such a strong gauge dependence
is not seen for $\Psi_4$. 


\begin{figure}[ht]    
\centerline{\resizebox{9cm}{!}{\includegraphics{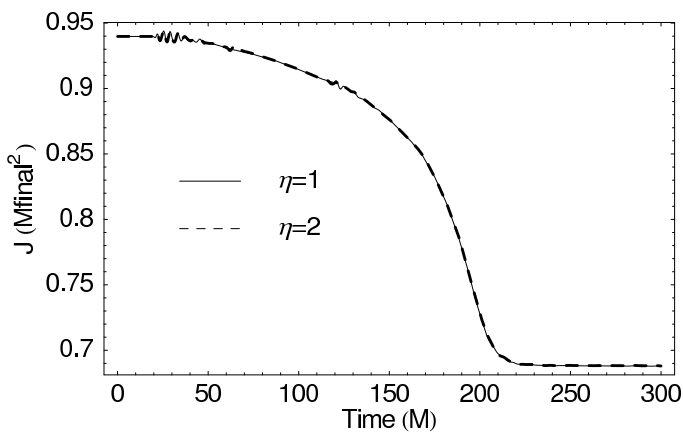}}}
\caption{Angular momentum computed at $r=30$ for gauge parameters $\eta=1,2$
with the BAM$\chi$ series --- no significant dependency on the gauge is seen.}
\label{fig:BondiJ}
\end{figure}            

\begin{figure}[ht]
\centerline{\resizebox{9cm}{!}{
\includegraphics{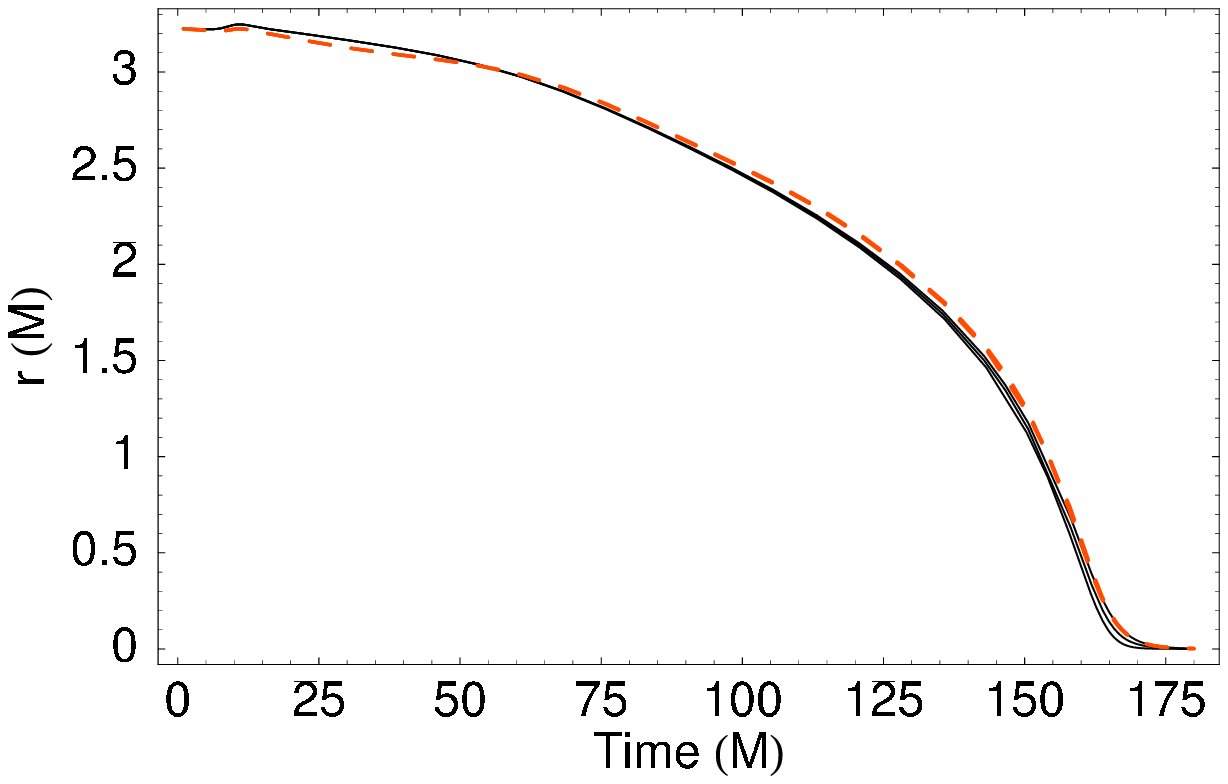}
\includegraphics{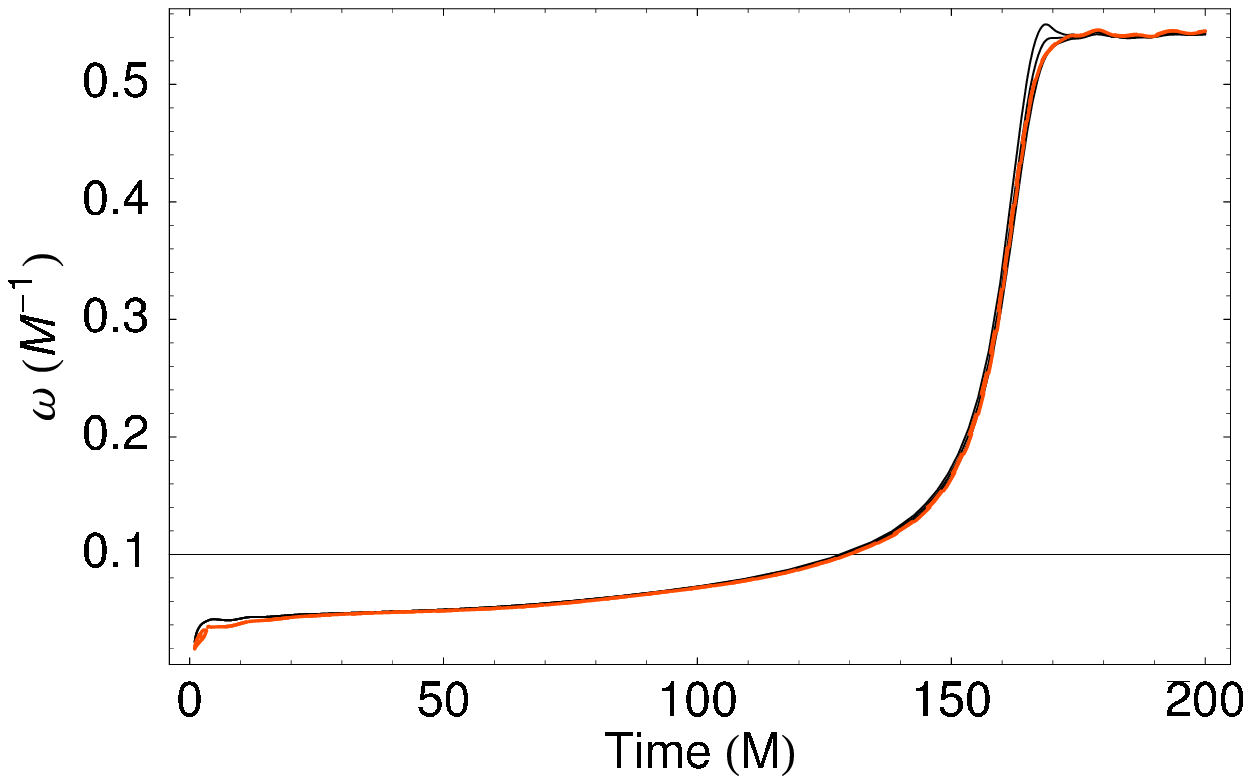}
}}
\centerline{
\includegraphics[width=6cm,height=3cm]{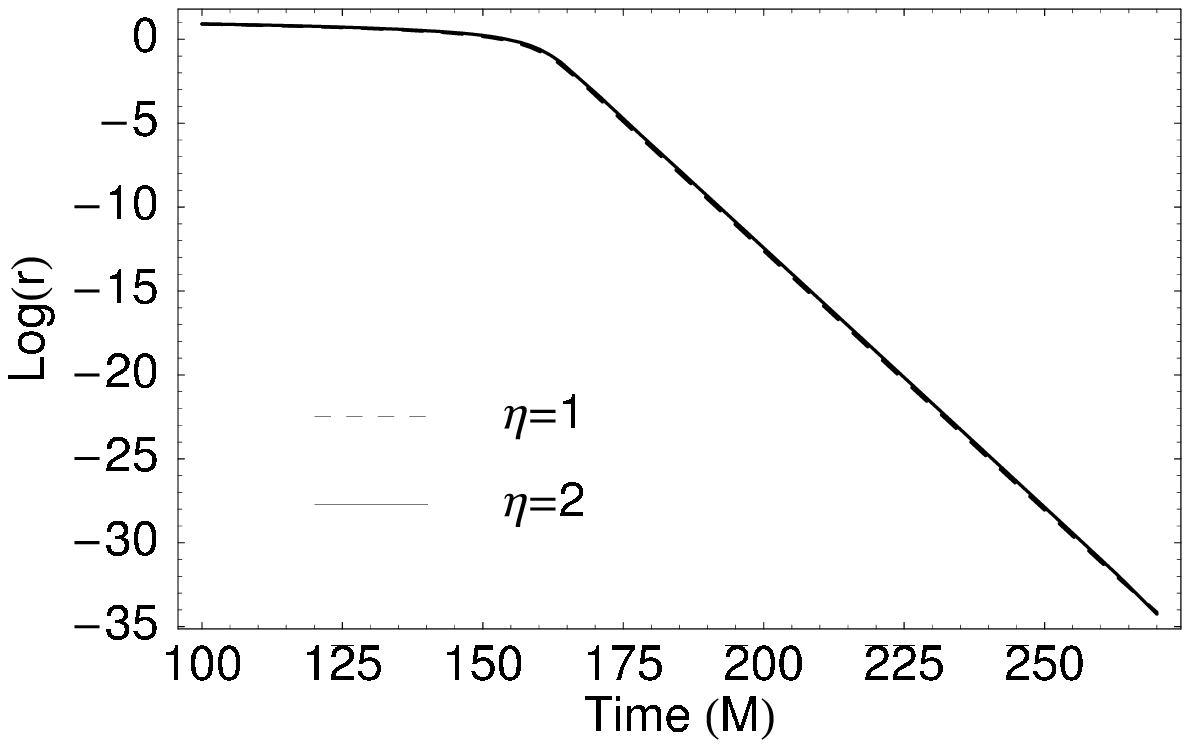}
}
\caption{Orbital motion with numerical errors
obtained from difference of Richardson extrapolated value to highest resolution result
(top left: r(t), top right: $\omega(t)$). The dashed curves represent  $\chi_{\eta=2}$ and the
full curves $\phi_{\eta=1}$.
Bottom: The logarithm of the puncture radial position
shows an exponential decay at late times, with a damping time of $\tau = 3.48 \pm 0.01$ ($M_{final}$).}
\label{fig:rw+errors}
\end{figure}

Important quantities to be determined are the amount of radiated angular
momentum, and the final spin of the black hole. For the time development of
the angular momentum as determined from the surface integral
Eq.~(\ref{jadm_int}) see Fig.~\ref{fig:BondiJ}. 
We determine the initial angular momentum by means of the 
surface integral Eq.~(\ref{jadm_int}), which can be evaluated analytically for
Bowen-York data as $J_0 = pD =0.866$, corresponding to a Kerr
parameter of $a/M_0 = 0.87$.  We numerically calculate the final angular
momentum as $J_{final} = 0.634$, corresponding to a Kerr parameter of
$a/M_{final} = 0.688$ and radiated angular momentum of $25\%$. The final
angular momentum has been estimated by several methods with an error of
roughly $1\%$. The surface integral Eq.~(\ref{jadm_int}) gives results that
are very accurate and consistent between different choices of the
$\eta$--parameter. We also examine the complex quasi-normal ringdown
frequency. The real and imaginary part of the quasi-normal ringdown frequency
can be determined by directly fitting an exponentially damped sinusoid to the
gravitational wave signal. The imaginary part (quality factor) can easily
be read off from $\vert \Psi_4\vert$, which shows exponential fall-off, and the
real part of the frequency can be determined from the time derivative of the
phase angle as defined in Eq.~(\ref{eq:phase_angle}). All these results are 
consistent, and yield a final spin of the black hole of $J = 0.683
M_{final}^2 \pm 1\%$, where  $M_{final}$ is the mass of the final black hole.  
Remarkably, the orbital frequency of the punctures levels off (see
Fig.~\ref{fig:rw+errors}) to the real part of the quasinormal frequency at
late times. Note that at early times, before the merger, the wave frequency
has been observed to be twice the orbital frequency \cite{Baker:2006yw}, as
would naively be expected from the quadrupole formula.


One of the remarkable facts about recent simulations of binary black holes,
whether done in a generalized harmonic or BSSN moving punctures framework, is 
the quality of the coordinate conditions: not only do they produce
a ``nice'' spiraling motion with almost spherical 
(apart from a short time during merger) apparent horizons as seen in Fig.\
\ref{fig:orbits_with_horizon}, but the coordinate tracks give rise to a good
estimate of the waves via the quadrupole formula (see, for example,
\cite{Buonanno:2006n}), and the measured
angular velocities coincide very accurately with what is expected on physical
grounds. For example, at the beginning of the simulation the angular velocity
quickly reaches a value close to that expected from the initial data
(approximately $M\Omega = 0.05$); see also \cite{Diener:2005mg}.
 A heuristic explanation for the latter fact has been given
in \cite{Hannam:2006vv}: symmetry-seeking gauge conditions (e.g., the
gamma-freezing condition used here) should be expected to find an approximate
helical Killing vector. This Killing vector will be unique up
to a rigid rotation of the form $\vec{\omega} \times \vec{r}$.
We can choose either corotation, or vanishing rotation
via the shift boundary condition at infinity. 
In this work the shift is set to zero at infinity. Thus the punctures' 
coordinate speeds are expected to be equal to their physical speeds 
seen from infinity.
It is interesting to check whether the choice of gauge -- in our approach this
boils down to the choice of shift damping parameter $\eta$ --- has an effect
on the coordinate tracks of the punctures. We plot the radial and angular
motion of the punctures with numerical errors (again determined from the
difference of the Richardson extrapolated value to highest resolution result)
for the BAM$\chi$2 and BAM$\phi$1 runs in Fig.\ \ref{fig:rw+errors}. While the
results look consistent between different runs, a small but significant
$\eta$-dependent deviation can be seen in the radial motion, whereas a
difference in angular motion is not visible on the plot. 

Fourth-order convergence of the puncture motion in the $\chi$2 and $\phi$1 runs is
demonstrated in Figs.~\ref{fig:r_convergence}, \ref{fig:w_convergence_chi2} and
\ref{fig:w_convergence_phi1}. The coordinate angular speed $\omega$ is calculated 
using $\omega = |\beta|_i/r$, where $|\beta|_i$ is the norm of the shift vector evaluated at 
the $i$th puncture that gives the coordinate speed of the puncture across the grid. This quantity shows 
fourth-order convergence up to the black-hole merger. After that time 
the puncture distance from the origin decays exponentially (see 
Fig.~\ref{fig:rw+errors}), thus rather soon the punctures are both less
than one grid point from the origin and the convergence in $\omega$ deteriorates to first-order.

\begin{figure}[ht]
\centerline{\resizebox{9cm}{!}{
\includegraphics{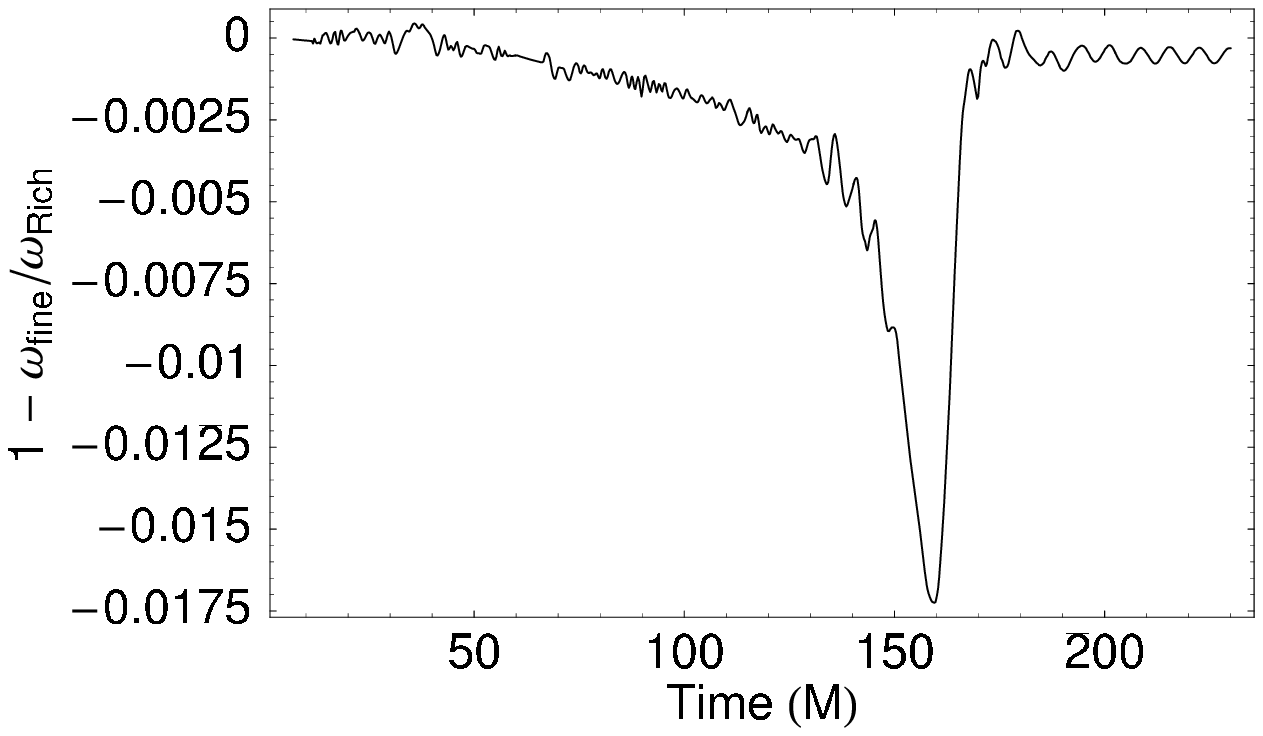}
\includegraphics{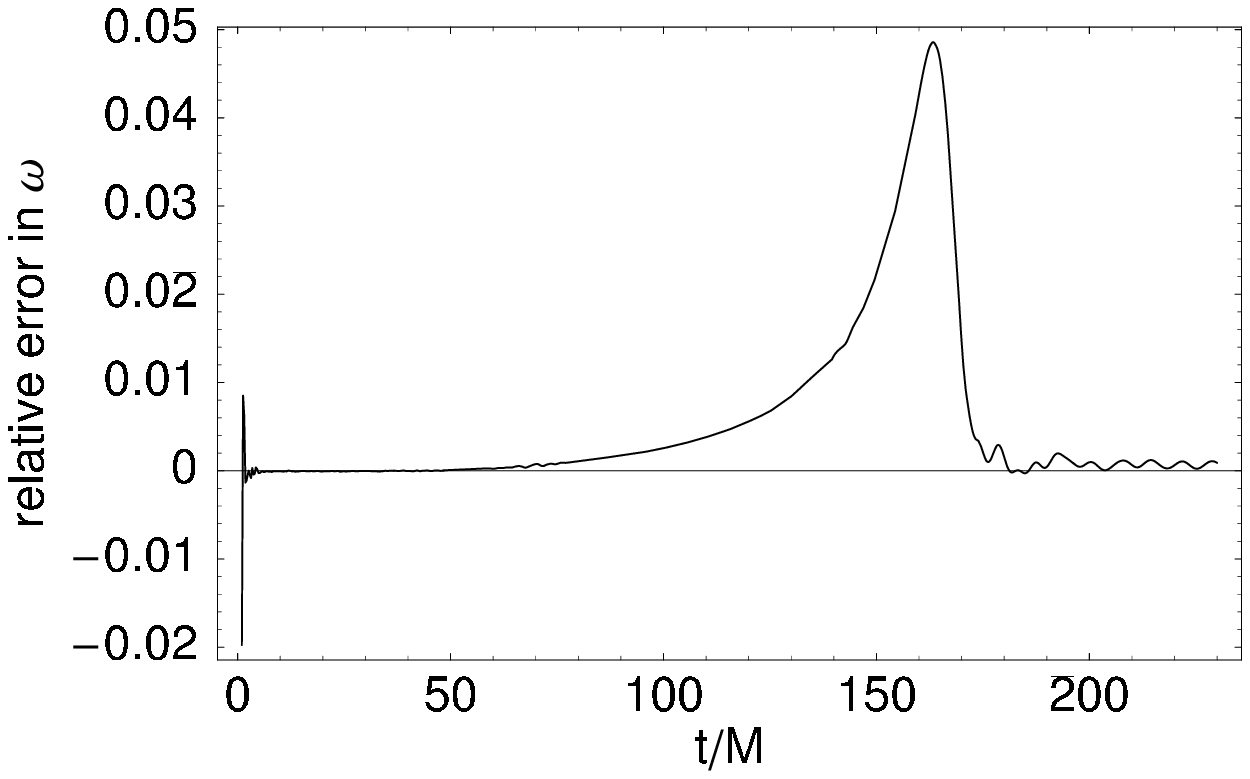}}}
\caption{Relative errors obtained from
difference of Richardson extrapolated value to highest resolution result
(left:  $\chi_{\eta=2}$, right: $\phi_{\eta=1}$).}
\label{fig:missing}
\end{figure}
%


\section{Quasi-circular orbit parameters}\label{sec:parameters}

For the runs in Section \ref{sec:Orbits} we chose the same initial parameters
as used by Baker, {\it et al.}~\cite{Baker:2006yw}, for their calibration
runs. These are in turn based on those from Cook's 1994 initial-data study
\cite{Cook94}. In that work the parameters for quasi-circular orbits were
determined using an ``effective potential'' (EP) method, whereby quasi-circular
orbits corresponding to a given total orbital angular momentum $J$ were
identified by the minimum in a curve of the binding energy $E_b = E_{ADM} -
M_1 - M_2$ versus the proper separation, in analogy with Newtonian
physics. Cook's results applied to inversion-symmetric (not puncture)
Bowen-York data, but a later study of the innermost stable circular orbit
(ISCO) of puncture Bowen-York data by Baumgarte \cite{Baumgarte00a} suggested
that there are only minor (if any) physical differences between the two types
of initial-data set.  

Note that Cook's parameters cannot be directly used
for punctures. For this reason, the parameters we actually use come
from Baker {\it et~al.}~\cite{Baker:2002qf} who have
translated Cook's parameters into puncture parameters.
As pointed out in~\cite{Tichy03a}, this translation has
introduced additional errors on the order of $1\%$ in the masses.
The translated Cook parameters have been found in practice to produce
reasonably convincing quasi-circular orbits. (However, see
\cite{Baker:2006yw,Buonanno:2006n} for evidence and estimates of eccentricity
for larger initial black-hole separations.) In 
addition, Baker {\it et al.}\ found that the merger waveform was largely
independent of the initial separation of the punctures, suggesting that the
parameters predicted by the effective-potential method really do correspond
to points on an inspiral sequence.  

However, there are a number of alternative ways to estimate the momenta as a
function of separation for black holes on an inspiral sequence. One option is
to use an approach based on the assumption of the existence of a helical Killing
vector (HKV) \cite{Gourgoulhon02,Grandclement02,Tichy03a}, 
for which a sequence of parameters for puncture data has been 
computed \cite{Tichy:2003qi}. Another option is to
use parameters predicted from post-Newtonian (PN) theory. How sensitive is the
final waveform to each of these approaches? 

Let us first consider post-Newtonian parameters. For a given separation $D$,
the momentum of each puncture can be given to 3PN order in the ADMTT gauge by
\cite{Damour:1999cr} \bea 
\frac{p}{\mu} & = &  \sqrt{\frac{M}{D}} +2 \epsilon
  \left(\frac{M}{D}\right)^{3/2} +\frac{1}{16} \epsilon ^2 \left(42- 43\nu \right)
\left(\frac{M}{D}\right)^{5/2}  \nonumber \\
& &  + \frac{ \epsilon ^3}{128} \left[ 480 +\left(163 \pi ^2 -4556  \right) \nu  + 104 \nu^2  \right]
   \left(\frac{M}{D}\right)^{7/2} .\nonumber \\
 \label{3PNp}
\eea The total mass is $M = M_1 + M_2$, the reduced mass is $\mu = M_1 M_2 / M$, $\nu =
\mu/M$, and the PN order of each term is indicated by $\epsilon$. For
equal-mass black holes with $M_1 = M_2 = 0.5$, we have $\mu = \nu = 0.25$. 

Eq.~(\ref{3PNp}) was derived using equations (5.1)-(5.3) in
\cite{Damour:1999cr}, and noting that $\omega_{static} = 0$ and
$\omega_{kinetic} = 41/24$ \cite{Damour2000_Poinc}. Equations (5.1) and (5.3)
can be rearranged, order by order in the post-Newtonian expansion, to give the
orbital angular momentum $J$ as a function of puncture separation $D$, and we
then use the relation $J = pD$ (which holds by definition; see section IV of
\cite{Damour:1999cr}) to write $p$ as a function of $D$. Eq.~(\ref{3PNp}) is
not gauge invariant, but the ADMTT gauge is expected to be 
very close to the conformally flat gauge that we choose for our initial data.  

\begin{figure}[t]
\centerline{\resizebox{8.5cm}{!}{\includegraphics[keepaspectratio]{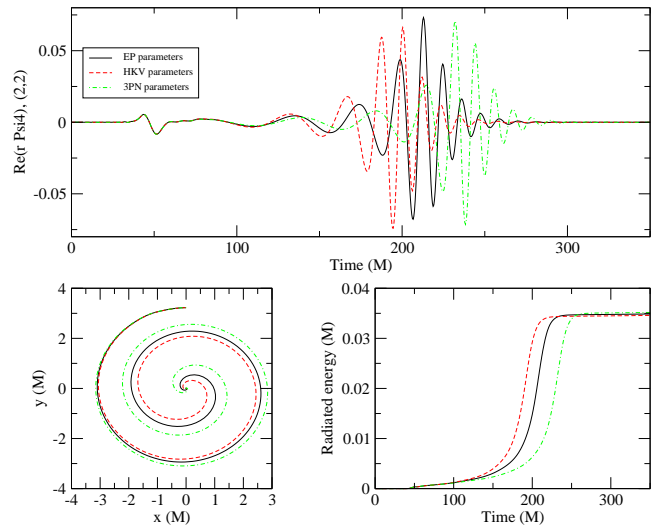}}} 
\caption{Results from evolution of Bowen-York puncture data using three
  choices of initial parameters, described in the text. Top: real part of
  the $l=2, m=2$ mode of $r \Psi_4$, extracted at $r = 40M$. Lower left: paths
  followed by the punctures during evolution. Lower right: energy
  extracted at $r=40M$ after $300M$ of evolution.}
\label{fig:comparison}
\end{figure}

Initial parameters from these three approaches, effective-potential (EP),
helical Killing vector (HKV), and third-order post-Newtonian (3PN), are given
in Table \ref{tab:parameters_BH}.
The parameters are scaled with respect to the total black-hole mass, $M
= M_1 + M_2$, which is a convenient quantity in all three approaches; in the
post-Newtonian expression (\ref{3PNp}) the black-hole masses appear, but the
other standard  mass scale, the total ADM mass $M_{ADM}$, does not. In
each case, the coordinate separation of the punctures is kept fixed, and a
prediction for the momenta that will produce a quasi-circular orbit is
provided. These predictions all differ by less than 1\% (which is also the
error estimate in Cook's sequence \cite{Cook94}), and we might expect the
resulting orbital motion and merger waveforms to be equally close. However,
from Figure \ref{fig:comparison} it is clear that this is not the case.

\begin{table}
\begin{ruledtabular}
\begin{tabular}{l|r|r|r}
Method                               & $m/M$  & $y/M$  & $p/M$  \\[10pt]
\hline
Effective-potential (EP)      & 0.4782 & 3.2248 & 0.1317   \\
Helical Killing vector (HKV)   & 0.4782 & 3.2248 & 0.1307   \\  
3PN                           & 0.4780 & 3.2248 & 0.1329  
\end{tabular}
\end{ruledtabular}
\caption{\label{tab:parameters_BH}Initial parameters for a given initial
  coordinate separation, from three different approaches: the
  effective-potential (EP) and helical Killing vector (HKV) methods, and the
  3PN formula (\ref{3PNp}). The parameters are scaled with respect to the
  total black-hole mass in the initial data.} 
\end{table}

The black-hole merger times differ by about $40M$, and the merger
waveforms are noticeably different. However, from the tracks of the puncture
locations during the evolutions, it is not clear which is the ``better''
choice of quasi-circular orbit parameters, or even 
what it would mean for one choice to be better than the other. These
evolutions also suggest that parameters calculated from PN methods are an
acceptable  alternative to numerically generated parameters, allowing a wide
range of configurations to be explored without the need for accompanying
initial-data studies. 

\begin{figure}[t]
\centerline{\resizebox{9cm}{!}
{
\includegraphics{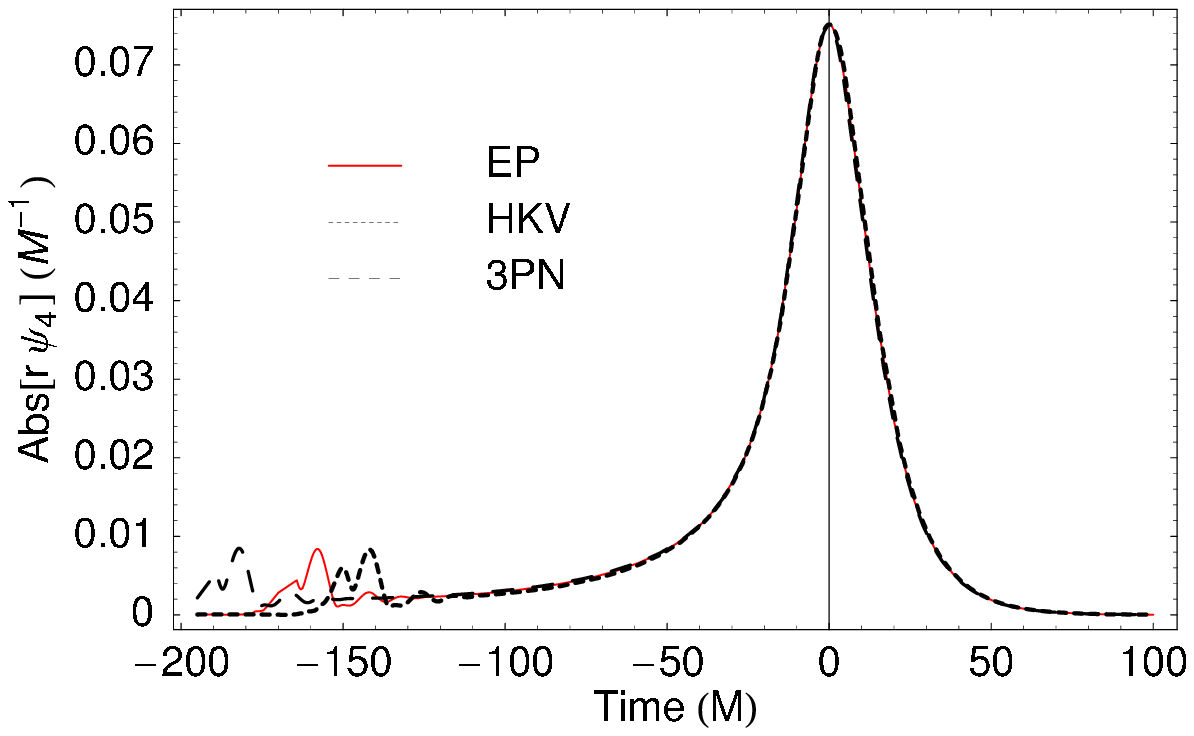}
\includegraphics{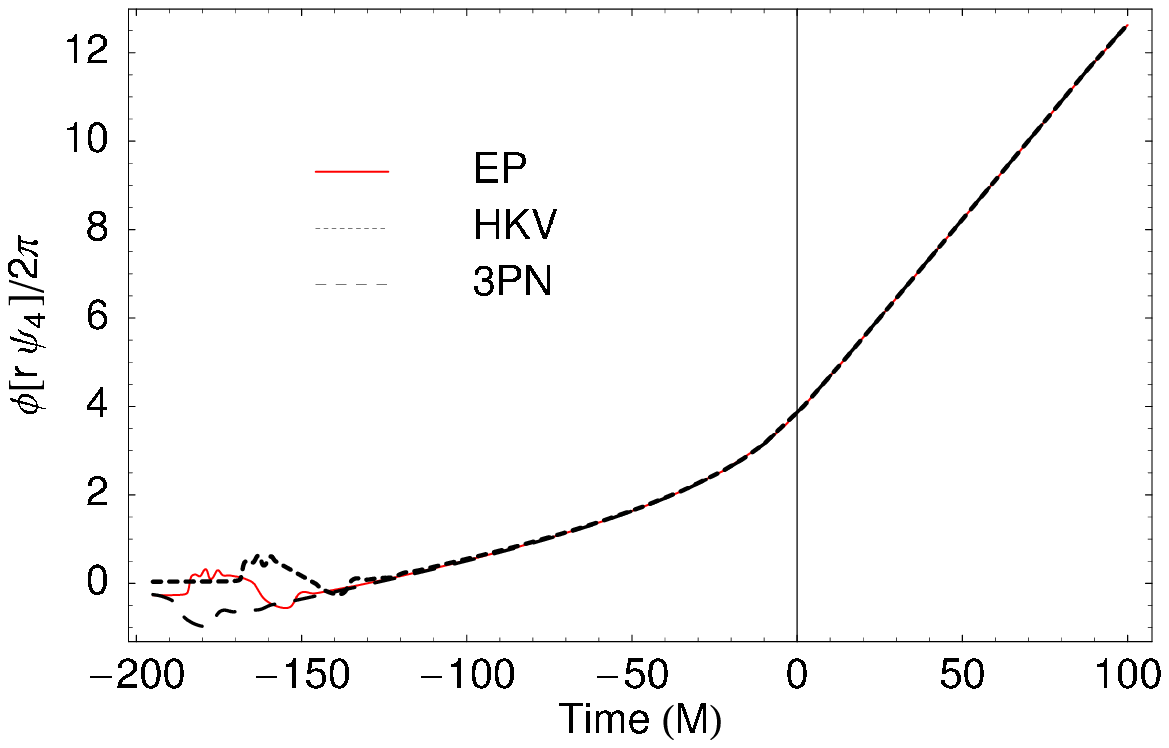}
}}
\caption{Results from evolution of Bowen-York puncture data using three
  choices of initial parameters, described in the text. Left: Absolute value of
  the $l=2, m=2$ mode of $r\Psi_4$, extracted at $r = 40M$. Right: 
  Phase angle (in units of $2\pi$, shifted in time to align the maxima
  of $abs(r\Psi_4)$ at $t=0$, and also aligning the phase at $t=0$.}
\label{fig:comparison2}
\end{figure}

Despite the apparent differences in the dynamics, the
radiated energy from the merger differs by only a few percent. In Figure
\ref{fig:comparison2} the amplitude and phase of $\Psi_4$ are shown
separately, with the results shifted in time so that the maximum in the
amplitude occurs at the same time for all three choices of initial parameters
(see \cite{Baker:2002qf,Baker:2006yw}). It is now clear that, although the
dynamics and waveforms look quite different, these differences are merely
cosmetic: the physics of the merger is the same for all three choices of
initial parameters.

\section{Discussion}\label{sec:Discussion}

We have presented a new code to evolve black-hole binaries, which is an
extension of the older BAM code
\cite{Bruegmann96,Bruegmann97,Bruegmann:2003aw}. The new BAM code implements
the ``moving puncture'' method (in both its $\phi$ and $\chi$ versions) within
a moving-box-based adaptive mesh-refinement grid structure, and uses
fourth-order-accurate spatial finite differencing and RK4 time evolution. The 
primary analysis tool, the extraction of gravitational radiation waveforms, is 
implemented using the $\Psi_4$ Newman-Penrose scalar.  

In this paper we have presented a number of important tests of our
code: evolutions of a single non-spinning black hole, a single
spinning black hole, and, the crucial test, evolutions of black-hole
binaries. In addition to demonstrating fourth-order convergence in regimes of
physical interest, each test has provided valuable insight into the grid
sizes, resolutions and geometries necessary to achieve accurate {\it
and} efficient simulations. One of the ultimate uses of our code will be
to perform large parameter studies of gravitational-wave sources, and
therefore efficiency of the code is of equal concern to its accuracy
and stability. The performance of the code, and how to tune its
configuration to obtain good performance, is one of our main results:
as shown in Table \ref{tab:timing} we are able to perform a
convergence series of three runs in the fourth-order convergent regime
at a total cost of roughly 1000 CPU hours and a maximal physical memory
consumption of less than 18 GByte.  Accurate results for binary black
hole evolutions can thus already be obtained with relatively small
commodity clusters, which are available in many research groups.

Evolutions of single black holes showed that both the $\phi$ and $\chi$
moving-puncture methods are stable and accurate, although the $\chi$ method
shows better convergence properties at the puncture. The moving-puncture
method was also found to be stable even when resolutions of up to $M/512$ were
used at the puncture. For black-hole binaries, the  $\chi$ method showed
monotonic convergence behavior (for example, in the merger times) at low
resolutions, and appears to us as clearly preferable, even though we have not
found prohibitive problems with the $\phi$ method.
As a further element of validating our code, we have
compared our waveform data with that from the
independent LEAN code \cite{Sperhake:2006cy}, and found that the two were in
excellent agreement. This is a strong validation of both codes, and it will be
interesting to see more extended comparisons between codes capable of
performing binary black hole simulations -- both regarding their results and
efficiency in order to further refine the methods of the field.

We have investigated the influence of gauge choice, and have found consistent
results between the 000 and ttt shift advection choices, as well as
the choice of the $\eta$ parameter in the $\tilde \Gamma$--driver shift
condition, e.g., also by comparing the LEAN and BAM results.
We have also found that, as expected, the $\eta$ parameter in the
$\tilde{\Gamma}$-driver shift condition has an effect on the final coordinates
of the solution, and as a result larger values of $\eta$ lead to a
larger coordinate size of the
black hole; in effect, the black hole becomes better resolved on the numerical
grid. This coordinate drift may however also cause problems
for naive wave extraction algorithms, and further research will be 
required to make optimal gauge choices.

Finally, we performed simulations using different choices of initial
parameters for the momenta of the punctures. We found that very small changes
in the initial momenta can make a large difference in the merger time of the
black holes, but do not change the physical properties of the radiation. 

Having carefully tested and calibrated our code for simulations of
comparable-mass black-hole binaries, in future publications we plan to
extend our research to parameter studies of unequal mass and spinning black holes, and
to use initial-data sets with larger separations.


\acknowledgments
It is a pleasure to thank Pedro Marronetti for discussions during the
course of this project; Bernard Kelly for access to his unpublished notes on
wave extraction; and Iris Christadler at the LRZ for assistance in code
optimization. 

This work was supported in part by 
DFG grant SFB/Transregio~7 ``Gravitational Wave Astronomy'' and by
NSF grant PHY-0555644. S. Husa acknowledges the hospitality of the
relativity group at the University of the Balearic Islands,
and has been supported in part by their CICYT grant FPA-2004-03666.
Computations where performed at HLRS, Stuttgart, and the LRZ,
Munich. We also acknowledge partial support by the National
Computational Science Alliance under grant PHY060040T.

\bibliography{references_cvs,references_extra}

\end{document}